\documentclass[a4paper,11pt]{article}

\usepackage{jheppub} 

\title{\boldmath Non-uniform black strings and the critical dimension in the $1/D$ expansion}

\author[a]{Ryotaku Suzuki}
\author[b]{and Kentaro Tanabe}

\affiliation[a]{Department of Physics, Osaka City University, Osaka 558-8585, Japan}
\affiliation[b]{Theory Center, Institute of Particles and Nuclear Studies, KEK,\\Tsukuba, Ibaraki, 305-0801, Japan}

\emailAdd{ryotaku@sci.osaka-cu.ac.jp}
\emailAdd{ktanabe@post.kek.jp}

\newcommand{\fr}[1]{\frac{1}{#1}}

\newcommand{\ord}[1]{{\mathcal O}(#1)}

\newcommand{\cA}{{\mathcal A}}
\newcommand{\cB}{{\mathcal B}}
\newcommand{\cC}{{\mathcal C}}

\newcommand{\cR}{{\mathcal R}}

\newcommand{\nonum}{\nonumber\\ }

\newcommand{\sR}{{\sf R}}
\newcommand{\sX}{x}
\newcommand{\sfa}{{\sf a}}
\newcommand{\sfb}{{\sf b}}

\newcommand{\tiM}{{\sf M}}
\newcommand{\tib}{b}

\newcommand{\farA}{{\delta \hspace{-1pt} A}}
\newcommand{\farB}{{\delta \hspace{-1pt} B}}
\newcommand{\farC}{{\delta \hspace{-1pt} C}}
\newcommand{\average}[1]{{\left\langle #1 \right\rangle}}
\newcommand{\cout}[1]{}
\newcommand{\Li}{{\rm Li}}

\newcommand{\omegaS}{\Omega}

\abstract{
Non-uniform black strings (NUBS) are studied by the large $D$ effective theory approach.
By solving the near-horizon geometry in the $1/D$ expansion, we obtain the effective equation for the deformed horizon up to the next-to-next-to-leading order (NNLO) in $1/D$. We also solve the far-zone geometry by the Newtonian approximation.
Matching the near and far zones, the thermodynamic variables are computed in the $1/D$ expansion.
As the result, the large $D$ analysis gives a critical dimension $D_*\simeq13.5$ at which the translation-symmetry-breaking phase transition changes between first and second order. This value of $D_*$ agrees perfectly, within the precision of the $1/D$ expansion, with the result previously obtained by E.~Sorkin through the numerical resolution. We also compare our NNLO results for the thermodynamics of NUBS to earlier numerical calculations, and find good agreement within the expected precision.}

\begin{document}
\rightline{KEK-TH-1827, AP-GR-125, OCU-PHYS-426} 
\maketitle

\section{Introduction}

Black holes in the dimension $D>4$ have totally different aspects than in $D=4$.
A remarkable property of higher dimensional black holes is the existence of dynamical instabilities
in the long wavelength,
which is originally discovered by Gregory and Laflamme for the uniform black string solution (UBS) whose horizon admits the isometry of  $S^{D-3}\times S^1$~\cite{Gregory:1993vy}.
The Gregory-Laflamme (GL) instability happens if the circumference of the compact dimension $L$ is in the same order of $r_0$, which is the radius of $S^{D-3}$.
It is also known that the zero mode of the GL instability generates another static branch, called
the non-uniform black string (NUBS), in which the translation symmetry in the compact dimension is broken.
NUBS solutions are constructed perturbatively~\cite{Gubser02,Sorkin04} and then numerically up to $D=15$~\cite{Kleihaus:2006ee,Wiseman:2002zc,Sorkin06,Headrick10,FiguerasMR12}.

The phase of NUBS for the fixed $L$ is known to admit a critical behavior with respect to the dimension~\cite{Sorkin04}.
In $D\leq 13$, NUBS becomes more massive than UBS at the GL critical point
and has less entropy than UBS with the same mass.
Therefore, the NUBS phase is unstable and cannot be the end point of the GL instability.
On the other hand, for $D>13$, NUBS becomes less massive and has more entropy, and then becomes stable.

Recently, the authors and other people have shown that, in the large number limit of the dimension $D\rightarrow \infty$, the black hole dynamics can be solved analytically by the expanding in the inverse dimension
 $1/D$~\cite{Asnin+07,EmparanST13,EmparanT14,Emparan14a,EmparanST14b,EmparanST15a}.
These simplification can be explain by the fact that the large $D$ limit of various black hole spacetimes  universally admit the $2$D string black hole structure, even in the existence of the cosmological constant and rotations~\cite{EmparanGT13}.

The key feature of the large $D$ limit is that the gravitational dynamics are divided into the two sectors
each with the separated scales in terms of the black hole radius $r_0$~\cite{Emparan14a}:
\begin{itemize}
\item Non-decoupling sector:  the dynamics with the gradient large as $\sim D/r_0$, which describe the degrees of freedom of the gravitational wave propagating to the asymptotic region.  
\item Decoupling sector: the dynamics with the smaller gradient ($\ll D/r_0$), confined within the short distance ($\sim r_0/D$) from the horizon, and decoupled from the asymptotic dynamics.
\end{itemize}
The dynamics reflecting the characteristics of each black hole, including its instabilities, is
described by the decoupling sector.
As for the GL instability of the black string, the $1/D$ expansion revealed that the threshold wavenumber approaches $k_{\rm GL} \simeq \sqrt{D}/r_0$ at $D\rightarrow \infty$, as suggested from the numerical result~\cite{Sorkin04}. Its correction up to the fourth order in $1/D$ was also calculated
and reproduced the numerical result, even for relatively lower dimensions $D\sim 10$~\cite{Asnin+07,EmparanST13,EmparanST15a}.

From the decoupling nature, 
one can expect that the dynamics of the small gradient degrees of freedom reduces to an effective theory living in the near-horizon geometry.
The authors and collaborators have shown that the Einstein equation for the decoupling sector is separated to the radial direction
with the large gradient $\partial_r \sim D/r_0$ and the dynamics along the horizon with the smaller gradient ($\ll D/r_0$)~\cite{EmparanSSTT15}.
Under this separation, the radial dependence can be easily integrated out by solving the ordinary differential equation, and the remaining 
dynamics along the horizon is governed by a simple effective equation, which is expressed as the `soap-flim' equation on the some constant surface of the radial coordinate placed at the short distance ($\sim r_0/D$) from the horizon. This formalism enables one to solve the Einstein equation even in the non-perturbative regime, provided the gradient along the horizon remains not so large ($\ll D/r_0$).
The authors of~\cite{Minwalla+15} also studied the large $D$ effective theory including the slow time dependence.

In this paper, we study the effective equation for NUBS up to the next-to-next-to-leading order (NNLO) in the $1/D$ expansion, extending the leading order result in~\cite{EmparanSSTT15}.
We also demonstrate the matching to the far-zone geometry and evaluate the asymptotic charges.
With these results, the thermodynamic variables are calculated up to NNLO.
The phase diagram including the NNLO correction admits the critical behavior around $D_*\simeq13.5$.\footnote{In the literature, the critical dimension $D_*$ is usually defined as the highest dimension with the first order transition: $D_*=13$. In this paper, we adopt the definition of $D_*$ as the real number at which the order of the transition changes, as in the condensed matter physics.
Because the large $D$ analysis treats the number $D$ just as a parameter, this is a rather natural definition.}
Within the error of $\ord{D^{-2}}$, this value agrees with numerical observation~\cite{Sorkin04,FiguerasMR12}.

The construction of this paper is as follows.
In section~\ref{sec:setup}, we explain the setup and outline of the analysis.
In section~\ref{sec:far}, the far-zone geometry is studied.
In section~\ref{sec:near}, the near zone is solved by the $1/D$ expansion and the effective equation up to NNLO is derived.
In section~\ref{sec:eft}, we study the effective equation.
In section~\ref{sec:phase}, the thermodynamics of NUBS is studied.
Section~\ref{sec:sum} gives the summary of this paper.
The appendices include the details of some lengthy equations used in the main part.
We attached a {\it Mathematica} file containing the detail of the NNLO near solution which are too lengthy to
show here. The file also contains the series solution for the effective equation.

\section{Setup}\label{sec:setup}
In this paper, we consider the $D=n+4$ static and spherically symmetric spacetime with one compact dimension. Such spacetimes are described by the following metric
\begin{eqnarray}
 ds^2 = -A(r,Z)dt^2+B(r,Z)(dr^2+dZ^2)+r^2C(r,Z)d\Omega_{n+1}^2,\label{eq:conformal-ansatz}
\end{eqnarray}
where we assume $A,B,C \rightarrow 1$ at $r\rightarrow \infty$ for the asymptotic flatness.
In this conformal ansatz, we solve the near-horizon geometry by the large D effective theory approach~\cite{EmparanSSTT15}. Because the gravity leaking from the near zone falls off by $1/r^n$, we rather use $1/n$ as the small expansion parameter, instead of $1/D$ itself.
For the near-zone analysis, we introduce the another radial coordinate suitable to describe the near-horizon structure with the large radial gradient ($\sim n/r_0$),
\begin{eqnarray}
 \sR = r^n/r_0^n.
\end{eqnarray}
We also introduce the new compact coordinate to capture the variation in the scale of $k_{\rm GL}\sim \sqrt{n}/r_0$,
\begin{eqnarray}
 Z = \frac{r_0 z}{\sqrt{n}}.\label{eq:z-def}
\end{eqnarray}
This scaling is a natural choice if we want to keep the self-gravitating effect finite
in the large $n$ limit.\footnote{In the earlier work~\cite{Camps:2010br,EmparanST13}, this scaling was also explained
by the hydrodynamical picture.} Assuming a bulge formed on the horizon with the wavelength $\lambda$ and radius deformed by $\delta r_0\sim r_0$, the mass excess from the original horizon
is roughly estimated by
\begin{eqnarray}
 G \delta M \sim \frac{2\pi^\frac{n+1}{2}r_0^n\lambda}{\Gamma\left(\frac{n+1}{2}\right)}.
\end{eqnarray}
Then, the extra gravitational potential made by this mass excess on its neighbor becomes
\begin{eqnarray}
  \delta \Phi \sim \frac{G\delta M}{\lambda^{n+1}} \sim \left(\frac{r_0\sqrt{2\pi e}}{n^{1/2}\lambda}\right)^n,
\end{eqnarray}
which means that the gravitational interaction between bulges formed at intervals of $\lambda$ has
the finite limit at $n\rightarrow \infty$ only if $\lambda \sim r_0/n^{1/2}$.
For more gentle deformations like $\lambda \sim r_0$, the deformed horizon needs other supporting forces in the limit.
For example, the authors and collaborators have demonstrated the negative cosmological constant
allows several deformed solutions: droplets and cavities~\cite{EmparanSSTT15},
and the rotation also allows bumpy Myers-Perry solutions~\cite{SuzukiTanabe15a}.
As for the Kaluza-Klein spacetime, we have localized black holes other than black strings~\cite{cagedbh,Headrick10},
in which case the gravitational forces from distant black holes support the deformation.
Though, such gravitational effect beyond the near zone will be rather contained in the non-decoupling sector with the large gradient.

In our scaling~(\ref{eq:z-def}), the static ansatz~(\ref{eq:conformal-ansatz}) allows the deformation only up to the magnitude of $r_0/n$ for the valid $1/n$ expansion~\cite{EmparanSSTT15}.
However, this does not mean the analysis is perturbative.
This magnitude of the deformation still produces the non-perturbative variation in the mass scale as $(r_0+\delta r_0(z)/n)^n \sim r_0^ne^{\delta r_0(z)/r_0}$. We will see that this non-perturbative degree of freedom follows a nonlinear effective equation.

\paragraph{Matching Strategy}
For the evaluation of the asymptotic charges such as the mass and string tension, the author and collaborators proposed the general way of matching with the far zone through the quasi-local energy momentum tensor~\cite{EmparanSSTT15}.
Instead, here we take the more primitive method.
We will directly match the near and far solution in the overlap region $1\ll {\sR} \ll e^n$.
The strategy is as follows.
The far solution is simply solved by the Newtonian approximation in the Minkowski background.
The Newtonian potential $\Phi(r,z)$ is expanded in the overlap region as
\begin{eqnarray}
 \Phi(r,z) \sim \frac{\phi(z)+\ord{\ln \sR/n}}{\sR}
\end{eqnarray}
where the function $\phi(z)$ is identified with the mass deformation $e^{\delta r_0(z)/r_0}$.
Since $\delta r_0(z)$ is only the functional degree of freedom in the near region, terms higher in $\ln \sR/n$ will not provide any additional information.
On the other hand, if we take $r\rightarrow \infty$, the Newtonian potential $\Phi(r,z)$ admits
the monopole behavior
\begin{eqnarray}
  \Phi(r,z) \sim \frac{\alpha r_0^{n}}{r^{n}}.
\end{eqnarray}
The important fact is that the monopole $\alpha$ is obtained by averaging $\phi(z)$ over $z$,
\begin{eqnarray}
 \alpha = \average{\phi} \sim \average{e^{\delta r_0(z)/r_0}}.
\end{eqnarray}
Since the total mass ($\sim \alpha r_0^n L$) will be the integration of the local mass density $e^{\delta r_0(z)/r_0}$, this is a natural consequence.
Therefore, once the near deformation $\delta r_0(z)$ is solved, the asymptotic charges are determined.

\section{Far zone analysis}\label{sec:far}
In this section, we examine the far-zone geometry, to which the near solution
in the $1/n$ expansion will be connected.
In the far zone $r-r_0 \gg r_0/n$, the gravitational potential $\sim r_0^n/r^n$
becomes much smaller than any power of $n$. 
Then, the far-zone geometry should be solved in the expansion of the Newtonian order $r_0^n/r^n$, instead of $1/n$.

We also study the boundary behavior of the far-zone geometry.
The far zone includes two boundary regions: {\it overlap} and {\it asymptotic} region.
In the overlap region, the far solution is further expanded by $1/n$ and matched with the near solution.
In the asymptotic region, we merely take $r\rightarrow \infty$ without taking the large n limit
to obtain the asymptotic charges.

\subsection{Newtonian approximation}
Let us perturb the metric~(\ref{eq:conformal-ansatz}) from the Minkowski background.\footnote{The similar analysis is found in~\cite{Harmark:2003eg}.}
\begin{eqnarray}
 ds^2 = - (1+\farA) dt^2 + (1+\farB)(dr^2+dZ^2)+r^2(1+\farC) d\Omega_{n+1}^2
\end{eqnarray}
where we write
\begin{eqnarray}
 A = 1+\farA,\quad B = 1+\farB,\quad C = 1+\farC\label{eq:far-newton-corr}
\end{eqnarray}
and each correction is assumed to be in the Newtonian order $\ord{r_0^n/r^n}$.
Then, the equation reduces to the Laplace equation in the cylindrically symmetric space.
\begin{eqnarray}
 \left( \partial_r^2 + \frac{n+1}{r}\partial_r + \partial_{Z}^2 \right)\Phi(r,Z)=0
\end{eqnarray}\cout{%
where $\Phi(r,Z)$ is introduced by
\begin{eqnarray},
 && \farA=-\Phi(r,Z),\quad
 \farB = - \frac{4\beta r_0^n}{(n+1)r^n} + \fr{n+1}\Phi(r,Z)\nonum
  &&\farC=\frac{8\beta r_0^n}{(n^2-1)r^n} + \fr{n+1} \Phi(r,Z) ,\label{eq:far-abc}
\end{eqnarray}}%
where $\Phi$ is introduced by
\begin{subequations}\label{eq:far-abc}
\begin{eqnarray}
 \farA=-\Phi(r,Z).
\end{eqnarray}
Other components reduce to
 \begin{align}
 \farB &= - \frac{4\beta r_0^n}{(n+1)r^n} + \frac{1}{n+1}\Phi(r,Z)+\partial_r\Psi(r,Z),\\
  \farC&=\frac{8\beta r_0^n}{(n^2-1)r^n} + \frac{1}{n+1}\Phi(r,Z)+r\Psi(r,Z),
\end{align}
\end{subequations}
where $\beta$ is an integration constant, and $\Psi(r,Z)$ follows
\begin{eqnarray}
( \partial_r^2 + \partial_Z^2)\Psi(r,Z) = 0.
\end{eqnarray}
One can realize that $\Psi(r,Z)$ is the degree of freedom of the conformal transformation. So we just set $\Psi(r,Z)=0$.
Assuming the periodicity $Z\rightarrow Z+ L$, the general solution
with the regular asymptotic behavior is given by the Fourier decomposition
\begin{eqnarray}
 \Phi(r,Z) =\frac{\alpha r_0^n}{r^n}+\sum_{j=1}^{\infty} \frac{a_j r_0^{n/2}}{  r^{n/2}} K_{n/2}(2\pi j r/L) \cos (2\pi j Z/L) \label{far-phi-sol}
\end{eqnarray}
where $K_{n/2}(x)$ is the modified Bessel function of the second kind.
The reflection symmetry $Z\rightarrow-Z$ is also imposed.
In the limit $r\rightarrow \infty$, only the first term can contribute to the asymptotic monopole, since $K_{n/2}(x)$ falls off exponentially at $x\rightarrow \infty$.

Here we note that eq.~(\ref{eq:far-abc}) admits the consistency relations
in the Newtonian order,
\begin{eqnarray}
 \delta A + 2 \delta B+ (n-1)\delta C = 0\label{eq:far-const-1}
\end{eqnarray}
and
\begin{eqnarray}
(n+1) \delta B + \delta A = - 4\beta/\sR \label{eq:far-const-2}.
\end{eqnarray}

\subsection{$1/n$ expansion in the overlap region}
In the overlap region, we introduce the rescaled wavenumber
\begin{eqnarray}
k=\frac{2\pi r_0}{\sqrt{n}L}.
\end{eqnarray}
With $r=r_0 \sR^{1/n}$,  the Debye expansion formula leads to
the expansion in terms of $1/n$,
\begin{eqnarray}
K_{n/2}( \sqrt{n} jk \sR^{1/n}) = \frac{f_j(k)}{\sqrt{\sR}} \left(1-\frac{j^2k^2}{n}\ln \sR+\frac{(-2j^2k^2+j^4k^4)\ln\sR+(j^4k^4/2-j^2k^2)\ln^2\sR}{n^2}\right)\nonum\label{eq:K-debye}
\end{eqnarray}
where $f_j(k)$ is defined up to NNLO in $1/n$ as
\begin{eqnarray}
&& f_j(k) = \left(1+\fr{6n}+\fr{72n^2}\right)\sqrt{\frac{\pi}{n}}\left(\frac{\sqrt{n}}{ejk}\right)^{n/2}e^{-\frac{j^2k^2}{2}}\nonum
&& \times \left(1+\frac{j^4k^4-4j^2k^2}{4n}+\fr{n^2}\left(-2j^2k^2+\frac{5j^4k^4}{2}-\frac{7j^6k^6}{12}+\frac{j^8k^8}{32}\right)\right).
\end{eqnarray}
This expansion will be rather formal which breaks down for the large $j$.
But we believe such high frequency components are negligible in the small gradient assumption.
Then, the $1/n$ expansion of eq.~(\ref{far-phi-sol}) becomes
\begin{eqnarray}
 \Phi= \fr{\sR}\left(\alpha+\sum_{j=1}^\infty f_j(k) a_j(k) \cos(j k z) +\ord{\ln \sR/n}\right),\label{eq:phi-debye}
\end{eqnarray}
where we also inserted $Z=r_0 z/\sqrt{n}$.
The leading behavior can be sumed to a functional degree of freedom
\begin{eqnarray}
 \phi(z) = \alpha+\sum_{j=1}^\infty f_j(k) a_j(k)\cos(j k z). \label{eq:def-phiz}
\end{eqnarray}
With this function, eq.~(\ref{eq:phi-debye}) up to the relevant order can be rewritten in this form
\begin{eqnarray}
 \Phi = \fr{\sR}\left(\phi(z)+\Phi^{(1)}(z)\frac{\ln  \sR}{n}+\Phi^{(2)}(z)\frac{\ln ^2 \sR}{n^2} +\ord{n^{-3}}\right) \label{eq:far-phi-expand},
\end{eqnarray}
where each $\Phi^{(i)}$ is given by
\begin{eqnarray}
 \Phi^{(1)}(z) = \phi''(z)+\fr{n}(2\phi''(z)+\phi^{(4)}(z)),\quad \Phi^{(2)}(z) = \phi''(z)+\fr{2}\phi^{(4)}(z).\label{eq:far-const-3}
\end{eqnarray}
We replaced the factor $j^2k^2$ in eq.~(\ref{eq:K-debye}) with $-\partial_z^2$.
Therefore, the far solution~(\ref{eq:far-abc}) is determined only by $\phi(z)$ and $\beta$ in the overlap region.
In the following, we see that $\beta$ is related to the tension.

\subsection{Asymptotic charges}
Once $\phi(z)$ and $\beta$ are specified, the asymptotic charges are calculated as follows.
In the asymptotic region $r \rightarrow \infty$, the leading behavior of $\Phi$ in eq.~(\ref{far-phi-sol}) is given by $\alpha$, which is equal to the average of $\phi(z)$ over $z$
\begin{eqnarray}
 \alpha = \average{\phi} \equiv \fr{L}\int_0^L \phi(z) dZ = \frac{k}{2\pi} \int_0^{2\pi/k} \phi(z) dz.
\end{eqnarray}
The far solution~(\ref{eq:far-abc}) gives the monopole at $r\rightarrow \infty$,
\begin{eqnarray}
   \delta A =  -\frac{\alpha r_0^n}{r^n},\quad \delta B  = \frac{\alpha-4\beta}{n+1}\frac{r_0^n}{r^n},\quad \delta C = \left(\frac{8\beta}{n^2-1}+\frac{\alpha}{n+1}\right)\frac{r_0^n}{r^n}. 
\end{eqnarray}
Therefore, the ADM formula determines the mass and tension as~\cite{HarmarkObers04b}
\begin{eqnarray}
&&  {\mathcal M} = \frac{\omegaS_{n+1}r_0^n L}{16\pi G} \frac{n(n+2)\average{\phi}+4\beta}{n+1},
\quad {\mathcal T} = \frac{\omegaS_{n+1}r_0^n}{4\pi G}\beta,\label{eq:ADM-mass-tension}
\end{eqnarray}
where $\omegaS_{n+1}=2\pi^{n/2+1}/\Gamma(n/2+1)$ is the volume of $S^{n+1}$. 

\section{Near zone analysis}\label{sec:near}
In this section, we study the near-horizon geometry in the $1/n$ expansion, following the previous formulation~\cite{EmparanSSTT15} up to the one higher order.
\subsection{Setup}
The near-horizon geometry can be separated into the radial sector with the large gradient ($\sim n/r_0$) and the other sector with the smaller gradient ($\ll n/r_0$),
\begin{eqnarray}
 ds^2 = N^2(\sR,z) \frac{d\sR^2}{n^2} + g_{\mu\nu} dx^\mu dx^\nu.
\end{eqnarray}
Under this assumption, it is convenient to decompose the Einstein equation as
\begin{eqnarray}
&& K^2-K^\mu{}_\nu K^\nu{}_\mu-R=0\label{eq:einK-1}\\
&& \nabla_\nu K^\nu{}_\mu - \nabla_\mu K = 0\label{eq:einK-2}\\
&& \frac{n}{N}\partial_{\sR} K^\mu{}_\nu = - K K^\mu{}_\nu + R^\mu{}_\nu - \fr{N}\nabla^\mu \nabla_\nu N\label{eq:einK-3}
\end{eqnarray}
where $R^\mu{}_\nu$ is the intrinsic curvature for $g_{\mu\nu}$ and
\begin{eqnarray}
 K^\mu{}_\nu = \frac{n}{2N} g^{\mu\sigma}\partial_{\sR}g_{\sigma\nu}.\label{eq:einK-def}
\end{eqnarray}
Our coordinate ansatz~(\ref{eq:conformal-ansatz}) leads to the intrinsic metric as
\begin{eqnarray}
 g_{\mu\nu} dx^\mu dx^\nu =  -A(\sR,z)dt^2 + B(\sR,z)\frac{r_0^2 dz^2}{n^2}+ r_0^2  \sR^{2/n}C(\sR,z) d\Omega_{n+1}^2,
\end{eqnarray}
with the conformal gauge
\begin{eqnarray}
 N(\sR,z) = \frac{r_0 B(\sR,z)}{\sR^\frac{n-1}{n}}.
\end{eqnarray}
Since the far solution~(\ref{eq:far-abc}) has only the normalizable corrections ($\sim 1/\sR$)
 in the overlap region, the boundary condition for $\sR\rightarrow \infty$ is simply
\begin{eqnarray}
 A = 1 + \ord{\sR^{-1}},\quad B = 1 + \ord{\sR^{-1}},\quad C= 1+ \ord{\sR^{-1}} \label{eq:near-bdry}.
\end{eqnarray}
The large dimensionality of $S^{n+1}$ and the rescaling~(\ref{eq:z-def}) also requires
\begin{eqnarray}
 B = 1 + \ord{n^{-1}},\quad C=1+\ord{n^{-1}}
\end{eqnarray}
for the valid $1/n$ expansion.\footnote{The validity of the expansion itself
requires only the constancy of the leading behavior in $B$ and $C$~\cite{EmparanSSTT15}.}
Hence, the metric components are expanded by $1/n$ as
\begin{eqnarray}
 A(\sR,z) = \sum_{k=0} \frac{A^{(k)}(\sR,z)}{n^k},\   B(\sR,z) =1+ \sum_{k=0} \frac{B^{(k)}(\sR,z)}{n^{k+1}},\  C(\sR,z) = 1+\sum_{k=0} \frac{C^{(k)}(\sR,z)}{n^{k+1}}.\nonum
\end{eqnarray}

\subsection{Leading order}
As was done in~\cite{EmparanSSTT15}, the leading equation for the mean curvature $K$
becomes
\begin{eqnarray}
 \frac{n\sR}{r_0} \partial_\sR K(\sR,z) = - K(\sR,z)^2+\frac{n^2}{r_0^2}.
\end{eqnarray}
If we expand $K$ as
\begin{eqnarray}
K(\sR,z) = \frac{n}{r_0} \sum_{k=0} \frac{K^{(k)}(\sR,z)}{n^{k}},
\end{eqnarray}
the leading solution $K^{(0)}$ is given by
\begin{eqnarray}
 K^{(0)} = -\frac{\sR^2+M(z)^2}{\sR^2-M(z)^2} \label{eq:leading-sol-K}
\end{eqnarray}
where $M(z)$ is an integration function denoting the horizon position.
Solving eqs.~(\ref{eq:einK-1}) and (\ref{eq:einK-3}), we obtain the leading solution
\begin{subequations}\label{eq:LO-sol}
\begin{eqnarray}
 A^{(0)}  =\left(\frac{\sR-M(z)}{\sR+M(z)}\right)^2,\quad C^{(0)} =  4 \ln\left(1+\frac{M(z)}{\sR}\right)\label{eq:LO-sol-AC}
\end{eqnarray}
and
\begin{eqnarray}
B^{(0)}=\frac{4(M'(z)^2-M(z)M''(z))}{M(z)^2}\ln \left(1+\frac{M(z)}{\sR}\right).\label{eq:LO-sol-B}
\end{eqnarray}
\end{subequations}
Adding to the condition~(\ref{eq:near-bdry}), the regularity at $\sR=M(z)$ is also imposed for $B$ and $C$. 
The vector constraint~(\ref{eq:einK-2}) is trivially satisfied in this order.
By setting $M(z)=1$, this solution reproduces the uniform black string metric~(\ref{eq:ubs-sol-LO}).
Here we note that $M(z)$ does not directly denote the deformation in the radius $\delta r_0(z)$, but
the variation of the mass scale $M(z) \sim e^{\delta r_0(z)/r_0}$.

Recalling the definition of $\sR$, the radius of the horizon embedding surface is given by
\begin{eqnarray}
 r_h(z) = r_0 M(z)^{1/n} = r_0 \exp\left(\fr{n}\ln M(z)\right).
\end{eqnarray}
Since we assumed $\delta r_h \sim \ord{r_0 /n}$, the expansion is valid only if $|\ln M(z)|\ll n$.

\paragraph{Gauge choice}
As another gauge choice, one can set $M(z)=1$ by the radial rescaling $\sR \rightarrow M(z)\sR$ and instead, use the integration function of $C^{(0)}$ as the degree of freedom~\cite{EmparanSSTT15},
\begin{eqnarray}
 C^{(0)} = C_0(z) + 4\ln\left(1+\fr{\sR}\right).
\end{eqnarray}
This choice makes the radial coordinate fit the equipotential surface, which leads to the simpler near horizon analysis.\footnote{In the current system, the similar coordinate was used by Harmark and Obers~\cite{HarmarkObers02}.}
In this `equipotential' gauge, the near deformation function enters in the non-normalizable behavior in the overlap region, $C^{(0)} = C_0(z)+\ord{\sR^{-1}}$. So, in exchange for the simplified near-zone analysis, the matching with the far-zone geometry requires a nontrivial coordinate transformation, involving both $\sR$ and $z$.

Instead, the current embedding condition~(\ref{eq:near-bdry}) makes the near coordinate rather fit the asymptotic geometry. 
The deformation in the non-normalizable behavior $C_0(z)$ is now absorbed to the horizon shift $M(z)$.
This choice is convenient to keep contact with the asymptotic coordinate, and in which the matching between the near and far zone will not require any further coordinate transformation other than $\sR = r^n/r_0^n$.\footnote{This only works if the embedding has the trivial leading behavior: $r=r_0+\ord{n^{-1}}$.
For the nontrivial embedding like $r=\cR(z)+\ord{n^{-1}}$ as considered in~\cite{EmparanSSTT15},
 the use of $M(z)$ will not simplify the matching.}
 
\subsection{Next-to-leading order}
The solution at NLO is similarly calculated. The horizon position is fixed to $\sR=M(z)$ by the renormalization of $M(z)$. In the appendix.~\ref{sec:NLO-sol}, we show the detail of $A^{(1)}$, $B^{(1)}$ and $C^{(1)}$.
At this order, the vector constraint~(\ref{eq:einK-2}) admits the first nontrivial condition for $M(z)$,
\begin{eqnarray}
0 = M^{(3)}(z)+M'(z)-\frac{2M'(z)M''(z)}{M(z)}+\frac{M'(z)^3}{M(z)^2}.\label{eq:LO-eff-dif}
\end{eqnarray}
This is equivalent to the differentiated version of the effective equation obtained in~\cite{EmparanSSTT15}.\footnote{One can find that $M(z)$ corresponds to $e^{2{\cal P}(z)}$ in \cite{EmparanSSTT15}.}
As pointed out in~\cite{EmparanSSTT15},  eq~(\ref{eq:LO-eff-dif}) is the system equivalent to the undamped Toda oscillator~\cite{toda}.

 \paragraph{Effective equation from the matching}
Due to the assumption $\delta r_h(z) \sim r_0/n$, the vector constraint~(\ref{eq:einK-2}) provides only the one lower order effective equation.
This situation can be remedied by the use of the far-zone information. We can also derive the leading order effective equation only from the leading solution~(\ref{eq:LO-sol}), just by using the far consistency relation~(\ref{eq:far-const-2}).
The leading solution~(\ref{eq:LO-sol}) expanded up to $\ord{\sR^{-1}}$ is matched with the far Newtonian correction~(\ref{eq:far-abc}) as
\begin{eqnarray}
 \delta A =  - \frac{4M(z)}{\sR},\quad \delta B = \frac{4(M'(z)^2-M(z)M''(z))}{nM(z)\sR},\quad \delta C = \frac{4M(z)}{n\sR}.\label{eq:near-ex-LO}
\end{eqnarray}
Substituting this into eq.~(\ref{eq:far-const-2}), we obtain the leading order effective equation
\begin{eqnarray}
 M(z) -\frac{M'(z)^2}{M(z)}+M''(z) = \beta. \label{eq:LO-eff-const}
\end{eqnarray}
This is just the integrated form of eq.~(\ref{eq:LO-eff-dif}).

We note that this derivation is due to the fact that the far-zone equation is solved explicitly, which depends on the asymptotic structure.
More general asymptotics will not allow such simplification.

\subsection{Next-to-next-to-leading order}
Finally, we obtained the solution up to the next-to-next-to-leading order (NNLO).
Because of its lengthy form, the detail of the NNLO solution is given in the attached {\it Mathematica} file.
To this order, the vector constraint~(\ref{eq:einK-2}) gives\footnote{The relation between this equation and the soap-film equation in~\cite{EmparanSSTT15} is not trivial, because our radial coordinate is not
normal to the horizon. We need the transformation to the normal coordinate to find the relation.}
\begin{eqnarray}
&&0 = M^{(3)}(z)+M'(z)-\frac{2M'(z)M''(z)}{M(z)}+\frac{M'(z)^3}{M(z)^2}\nonum
&&-\fr{n}\left[(5-2\zeta(2)+4\ln 2)\left(\frac{M''(z)^2M'(z)}{M(z)^2}+\frac{M'(z)^5}{M(z)^4}\right)+(1+4\ln 2+2\ln  M(z))M'(z)\right.\nonum
&&\left.+\frac{(2\zeta(2)-2-8\ln 2)M'(z)^3}{M(z)^2}+\left(\frac{(8\ln2 -2\zeta(2))M'(z)}{M(z)}+\frac{(4\zeta(2)-10-8\ln 2)M'(z)^3}{M(z)^3}\right)M''(z)\right]\nonum \label{eq:mom-const-nnlo}
\end{eqnarray}
where the derivatives higher than $M''(z)$ in $\ord{n^{-1}}$ are eliminated by using this equation iteratively.
This effective equation determines the horizon deformation only up to NLO.
We will see that the one higher order equation can be obtained from the matching.

\subsection{Matching}
Now, we identify the Newtonian correction in the far zone~(\ref{eq:far-newton-corr}) with the near solution up to NNLO.
Expanding the near solution up to $\ord{\sR^{-1}}$, we obtain
\begin{subequations}
\begin{eqnarray}
 \sum_{i=0}^{2}\frac{A^{(i)}}{n^i}&=& 1-\fr{\sR} \left(\cA_0(z)+\cA_1(z)\ln  \sR/n+ \cA_2(z) (\ln  \sR/n)^2\right)+\ord{\sR^{-2}},\label{eq:near-expansion-overlap-1}\\
1+ \fr{n}\sum_{i=0}^{2}\frac{B^{(i)}}{n^i}&=& 1+\fr{n\sR} \left(\cB_0(z)+\cB_1(z) \ln  \sR/n+ \cB_2(z) (\ln  \sR/n)^2\right)+\ord{\sR^{-2}},\label{eq:near-expansion-overlap-2}\\
 1+\fr{n} \sum_{i=0}^{2}\frac{C^{(i)}}{n^i}&=& 1+\fr{n\sR} \left(\cC_0(z)+\cC_1(z)\ln  \sR/n+ \cC_2 (z)(\ln \sR/n)^2\right)+\ord{\sR^{-2}}\label{eq:near-expansion-overlap-3},
\end{eqnarray}\label{eq:near-expansion-overlap}
\end{subequations}
where each coefficient consists of $M(z)$ and its derivatives as in eq.~(\ref{eq:near-ex-LO}).
We show only the detail of $\cA_0$ in the appendix~\ref{sec:nnlomatch}.
The Newtonian correction $\delta A,\ \delta B,\ \delta C$ in the far zone~(\ref{eq:far-abc}) is matched with $\ord{\sR^{-1}}$ terms in the above equation.
As in the leading order, eq.~(\ref{eq:far-const-2}) for $\ord{(\ln \sR)^0/\sR}$ gives the effective equation up to NNLO
\begin{eqnarray}
 \cA_0(z)-(n+1)\cB_0(z)/n= 4\beta. \label{eq:far-const-0}
\end{eqnarray}
The detail of this equation is also shown in the appendix~\ref{sec:nnlomatch}.
The vector constraint~(\ref{eq:mom-const-nnlo}) is equivalent to the derivative of this equation up to $\ord{n^{-1}}$.
The match of $\delta A$ also determines $\phi(z)$ as
\begin{eqnarray}
 \phi(z) = \cA_0(z). \label{eq:NNLOmatch-phiz}
\end{eqnarray}
Therefore, we have completely determined the far-zone geometry up to NNLO in $1/n$.

\paragraph{Consistency check}
Eq.~(\ref{eq:near-expansion-overlap}) also contains the power of $\ln \sR/n$.
Since we have no more undetermined degree of freedom, the constraints~(\ref{eq:far-const-1}) and (\ref{eq:far-const-2}) should be trivially satisfied for such terms.
Actually, we found that our near solution satisfies up to the relevant order
\begin{eqnarray}
 \cA_k - 2\cB_k/n-(n-1)\cC_k/n =0\quad (k=0,1,2) \label{eq:near-const-1}
\end{eqnarray}
and
\begin{eqnarray}
 \cA_1-(n+1)\cB_1/n=0,\quad  \cA_2-(n+1)\cB_2/n =0.\label{eq:near-const-2}
\end{eqnarray}
Using eqs.~(\ref{eq:mom-const-nnlo}) and (\ref{eq:NNLOmatch-phiz}), the remaining consistency~(\ref{eq:far-const-3}) is also satisfied.

\section{Analysis of the effective equation}\label{sec:eft}
In this section, we study the non-uniform black string solution
by solving the NNLO effective equation~(\ref{eq:far-const-0}).
\subsection{Integrated form}
After an integration, eq.~(\ref{eq:far-const-0}) is arranged to the following form
\begin{eqnarray}
&&0=\frac{M'(z)^2}{2 M(z)^2}-\sfa_0-\sfa_1\ln  M(z)-\sfa_2\frac{\ln ^2 M(z)}{n}-\sfa_3\frac{\ln ^3 M(z)}{n^2}\nonum
&&+\ \frac{\beta}{M(z)}\left(\sfb_{0,0}+\sfb_{0,1}\frac{\ln  M(z)}{n}+\sfb_{0,2}\frac{\ln ^2 M(z)}{n^2}\right)
+\frac{\beta^2}{nM(z)^2}\left(\sfb_{1,0}+\sfb_{1,1}\frac{\ln  M(z)}{n}\right)
+\frac{\sfb_{2,0}\beta^3}{n^2M(z)^3}\nonum\label{eq:horizon-eq-2nd}
\end{eqnarray}
where $\beta$ is the rescaled value of tension in eq.~(\ref{eq:ADM-mass-tension}).
Each coefficient is given by
\begin{subequations}
\begin{eqnarray}
\begin{split}
\hspace{-0.5cm}\sfa_0&=1+a+\frac{1+4a+2a^2+4\ln2}{2n}+\frac{1}{n^2}\left(a+2a^2+\frac{2}{3}a^3+2(2a+1)\ln2\right),\\
\hspace{-0.5cm} \sfa_1 &= -1-\frac{4(a+1)}{n}-\frac{2(1+4a+2a^2+4\ln2)}{n^2},\quad \sfa_2 = 3+\frac{8(a+1)}{n},\quad \sfa_3 = -\frac{14}{3},
\end{split}
\end{eqnarray}
\begin{eqnarray}
\begin{split}
 \sfb_{0,0}&=1+\frac{3+4\ln  2-\zeta(2)-2 a(\zeta(2)-2)}{n}\\
 &\quad+\ \fr{10n^2}\left(30+80\ln ^2 2-80(\zeta(2)-2)\ln 2
 -70\zeta(2)+49\zeta(2)^2+4a^2\left(20-25\zeta(2)+7\zeta(2)^2\right)\right.\\
 &\qquad\qquad\left.-\ 15\zeta(3)-2a\left(-60+40(\zeta(2)-2)\ln  2+80\zeta(2)-42\zeta(2)^2+15\zeta(3)\right)\right),\\
 \sfb_{0,1}&=2\zeta(2)-6\\
 &\quad+\ \fr{n}\left(-18+8(\zeta(2)-3)\ln 2+18\zeta(2)-\frac{42}{5}\zeta(2)^2+3\zeta(3)-\frac{4a}{5}(30-30\zeta(2)+7\zeta(2)^2)\right),\\
 \sfb_{0,2}&=18-14\zeta(2)+\frac{14}{5}\zeta(2)^2,
\end{split}
\end{eqnarray}
\begin{eqnarray}
\begin{split}
 \sfb_{1,0}& =- \frac{5}{2}-2\ln 2+\zeta(2)-\fr{n}\left(10+8\ln ^2 2+(26-20\zeta(2))\ln 2-13\zeta(2)
 +\frac{46}{5}\zeta(2)^2\right. \\
 &\quad\left.+\ a\left(15-8\ln ^22-16\zeta(2)+\frac{44}{5}\zeta(2)^2-4(4\zeta(2)-3)\ln 2+2\zeta(3)\right)\right),\\
 \sfb_{1,1}&=20-8\ln ^22-16(\zeta(2)-1)\ln 2-18\zeta(2)+\frac{44}{5}\zeta(2)^2+2\zeta(3)
 \end{split}
\end{eqnarray}
and
 \begin{eqnarray}
 &&\sfb_{2,0} = 7-12(\zeta(2)-1)\ln 2-6\zeta(2)+\frac{24}{5}\zeta(2)^2+\frac{3}{2}\zeta(3).
\end{eqnarray}\end{subequations}
where $a$ is an integration constant, whose definition is tuned for the later convenience.
This is one of our main results. Here $\ln 2$ in the coefficients is just originating from the current conformal coordinate, and does not affect the physical quantities.

Eq.~(\ref{eq:horizon-eq-2nd}) is invariant under the rescaling,
\begin{eqnarray}
  M(z) \rightarrow M_1(z_1)=e^{\xi}M(e^{-\xi/n}z_1),\quad \beta \rightarrow \beta_1=e^{\xi} \beta,\quad a\rightarrow a_1 = a+\xi. \label{eq:rescaling-def}
\end{eqnarray}
The definition of $a$ is arranged so that this transformation works linearly on $a$.
Then, the constant $a$ is found to be the scaling degree of freedom of the spacetime,
 which can be set to an arbitrary value. The scaling at $a=0$ is also fixed so that $\beta=1$ just admits the uniform solution.

\subsection{Leading order equation}
Let us start with the leading order equation
\begin{eqnarray}
  \ln M + \frac{M'^2}{2M^2} +\frac{\beta}{M}-a-1=0\label{eq:LO-eq-int}.
\end{eqnarray}
For the time being, we put aside the scaling $a$ by setting $a=0$.
If we define $\varphi(z) = \sqrt{M(z)}$, then eq.~(\ref{eq:LO-eq-int}) reduces to the potential problem of the classical dynamics,
\begin{eqnarray}
 \fr{2}\varphi'^2 =- \frac{\beta}{4}-V(\varphi),\quad V(\varphi) = \fr{4}(\varphi^2\ln\varphi^2 -\varphi^2)
\end{eqnarray}
where the minus of the rescaled tension $-\beta$ plays the roll of the energy.
The minimum energy state with $\beta =1$ is given by the uniform solution $\varphi(z)=1$.
For the positive tension $0<\beta<1$, this potential also admits the oscillatory solution, which corresponds to the non-uniform black string (NUBS). Since the expansion breaks down at $M(z)=0$, the solution for the negative tension $\beta<0$
is not the physical branch. The solution also collapse to $M(z)=0$ for the zero tension $\beta=0$. However,  the zero tension solution can be identified with localized black holes, which will be discussed in the later section.

The maximum and minimum value for NUBS are given by
\begin{eqnarray}
 M_{\{\rm max,min\}}(\beta) = \frac{-\beta}{W_{\{0,-1\}}(-\beta/e)}\label{eq:maxminM}
\end{eqnarray}
where $W_{\{0,-1\}}(z)$ denote the upper and lower branches of the Lambert W function, defined as the inverse function for $z(w)=we^w$.
In the limit $\beta \rightarrow +0$, we have
\begin{eqnarray}
M_{\rm max} \simeq e,\quad M_{\rm min} \simeq \beta/\ln(1/\beta).\label{eq:maxminM-apr}
\end{eqnarray}
Since the expansion is only valid for $|\ln M(z)|\ll n$, the range of the parameters is restricted to $\beta \gg e^{-n}$.

\paragraph{Non-uniformity parameter}
An useful dimensionless measure for the deformation is the non-uniformity parameter~\cite{Gubser02} 
\begin{eqnarray}
\lambda = \fr{2}\left(\frac{R_{\rm max}}{R_{\rm min}}-1\right)
\end{eqnarray}
where $R_{\rm max}$ and $R_{\rm min}$ is the maximum and minimum radius of $S^{n+1}$.
This is expressed by the maximum and minimum value of $M(z)$,
\begin{eqnarray}
 \lambda = \fr{2}\left(\frac{M_{\rm max}(\beta)^{1/n}}{M_{\rm min}(\beta)^{1/n}}-1\right) .\label{eq:non-uniform-param-def}
 \end{eqnarray}
Using eq.~(\ref{eq:maxminM}),
 the non-uniformity parameter is estimated for small $\beta$ and large $n$ as
 \begin{eqnarray}
 \lambda \simeq \frac{\ln(1/\beta)}{2n} \ll \fr{2},
\end{eqnarray}
where we assumed $\beta \gg e^{-n}$.
Thus, the $1/D$ expansion covers the deformation only with the small non-uniformity $\lambda\ll \ord{1}$.
\cout{%
Using eq.~(\ref{eq:maxminM-apr}),
 the non-uniformity parameter is estimated for a small fixed value of $\beta$ and large $n$ as
 \begin{eqnarray}
 \lambda \simeq \frac{\ln(1/\beta)}{2n} 
\end{eqnarray}
which implies $\lambda \sim 1/2$ for $\beta \sim e^{-n}$ where the expansion becomes bad.
Thus, the $1/D$ expansion covers the deformation with the small non-uniformity $\lambda\ll \ord{1}$.
}%

\subsection{Higher order corrections}
Now, we study the effective equation including higher order corrections.
For $a=0$, eq.~(\ref{eq:horizon-eq-2nd}) is arranged to gives the UBS solution for $\beta=1$ as in the leading analysis,
\begin{eqnarray}
 M_{\rm UBS}= 1-\frac{\zeta(2)}{n}+\fr{2n^2}\left(\zeta(2)^2-4\zeta(3)\right).
\end{eqnarray}
The correction reproduces eq.~(\ref{eq:ubs-rh-r0}), if we identify $r_h = r_0 (M_{\rm UBS})^{1/n}$.
In the following analysis, we introduce the scale invariant variables $\tiM(z)$ and $\tib$ defined by
\begin{eqnarray}
 M(z) = e^{a} \tiM(e^{-a/n}z),\quad \beta = e^{a} \tib. \label{eq:scale-inv-def}
\end{eqnarray}
For $a=0$, $\tiM(z)$ and $\tib$ coincide with $M(z)$ and $\beta$. The boundary values for $\tiM(z)$ is given by
\begin{eqnarray}
&& \tiM_{\{\rm max,min\}}(\tib) = \frac{-\tib}{W_{\{0,-1\}}(-\tib/e)}\left[1+\fr{2n}\left(4\ln2-1+\left(2\zeta(2)-1+4 \ln2 \right)W_{\{0,-1\}}(\tib/e)\right)\right.\nonum
&&  +\  \fr{120n^2}\Bigl(5(11-72\ln2+48\ln^2 2) \nonum
&&\quad+\ (65+120(-5+4\zeta(2))\ln2+720\ln^22-240\zeta(2)-84\zeta(2)^2+180\zeta(3))W_{\{0,-1\}}(-\tib/e)\nonum
&&\quad \left.+\ 2(5+120(-1+2\zeta(2))\ln2+240\ln^22-120\zeta(2)-12\zeta(2)^2-30\zeta(3))W_{\{0,-1\}}(-\tib/e)^2\Bigr)\right].\nonum \label{eq:m-inv-maxmin}
\end{eqnarray}
One can see that, by the scaling property, the non-uniformity parameter~(\ref{eq:non-uniform-param-def}) depends only on the scale invariant parameter $\tib$,
\begin{eqnarray}
 \lambda = \fr{2}\left(\frac{\tiM_{\rm max}(\tib)^{1/n}}{\tiM_{\rm min}(\tib)^{1/n}}-1\right) .\label{eq:non-uniform-param-def-inv}
\end{eqnarray}

\paragraph{Near uniform solution}
Since we cannot take $\tib$ to be as small as $e^{-n}$, it suffices to consider the expansion around UBS, $\tib=1$.  Let us go back to the leading equation~(\ref{eq:LO-eq-int}). Expanding with $\tib=1-\epsilon^2/2$ and $\tiM(z) = 1+\delta \tiM(z)$, eq.~(\ref{eq:LO-eq-int}) admits the small oscillation in $\ord{\epsilon}$
\begin{eqnarray}
 \delta  \tiM(z) = \epsilon \cos(z)
\end{eqnarray}
where we assumed the reflection symmetry at $z=0$, $\tiM(z)=\tiM(-z)$.
Since the proper length is given by $Z=r_0 z/\sqrt{n}$, this reproduces the leading behavior of the GL threshold wavenumber $k_{\rm GL} \simeq \sqrt{n}/r_0$.
Due to the nonlinearity in eq.~(\ref{eq:LO-eq-int}), the period is also corrected
for the absence of the secular terms as $z \sin(z)$,
\begin{eqnarray}
 \cos(z) \rightarrow \cos (k(\epsilon) z).
\end{eqnarray}
The first correction to $k(\epsilon)$ is obtained at $\ord{\epsilon^2}$
\begin{eqnarray}
  k(\epsilon) = 1 - \frac{1}{24}\epsilon^2 + \ord{\epsilon^4},
\end{eqnarray}
which is determined by the solution up to $\ord{\epsilon^3}$,
\begin{eqnarray}
  \tiM(z) = 1+\epsilon \cos(k(\epsilon)z)+\frac{\epsilon^2}{6} \cos(2k(\epsilon)z)-\frac{\epsilon^3}{96}\left(\frac{7}{3}\cos(k(\epsilon)z)-\cos(3k(\epsilon)z)\right)+\ord{\epsilon^4}.\nonum
\end{eqnarray}

The inclusion of higher order corrections in eq.~(\ref{eq:horizon-eq-2nd}) is rather straightforward.
We just expand $\tiM$ by $1/n$ as $\tiM = \tiM_0+\tiM_1/n+\tiM_2/n^2$, and 
the avoidance of the secular terms also adds the higher order corrections to the wavelength $k(\epsilon) = k_0(\epsilon)+k_1(\epsilon)/n+k_2(\epsilon)/n^2$.
If we continue the analysis, the resulting solution can be written in the Fourier series
\begin{eqnarray}
 \tiM(z) = \sum_{j=0}^\infty \mu_j(\epsilon) \epsilon^j \cos(j k(\epsilon) z).\label{eq:nnlo-ser-sol}
\end{eqnarray}
where the $j$-th harmonics begins from $\ord{\epsilon^j}$.
We show only the first few terms in the appendix.~\ref{sec:nnloser}.
Further terms up to $\ord{\epsilon^7}$ are given in the {\it Mathematica} file attached to this paper.

The wavenumber $k(\epsilon)$ determined by the expansion up to $\ord{\epsilon^7}$ is given by
\begin{eqnarray}
k(\epsilon) &=& 4^{-1/n} \left(1-\fr{2n}+\frac{7}{8n^2}\right)\left[1-\left(1-\frac{9}{n}+\frac{7}{n^2}\right) \frac{\epsilon ^2}{24}
   -\left(\frac{7}{768}-\frac{9}{128 n}+\frac{3}{256 n^2}\right) \epsilon ^4\right. \nonum
&&  \hspace{3cm} \left.-\ \left(\frac{6971}{2488320}-\frac{379}{18432 n}-\frac{23}{41472 n^2}\right) \epsilon
   ^6+\ord{\epsilon^8}\right]. \label{eq:nnlo-kgl}
\end{eqnarray}
In the UBS limit $\epsilon \rightarrow 0$, $k(\epsilon)$ reproduces the higher order corrections to $k_{\rm GL}$ shown by the linear analysis~\cite{Asnin+07, EmparanST13,EmparanST15a}, except the factor $4^{-1/n}$. Actually, we will see that the mass length of the corresponding UBS is given by $\rho_0 = 4^{1/n}e^{a/n}r_0$. Then, if we recover the scaling~(\ref{eq:scale-inv-def}), the physical wavenumber at $\epsilon=0$ is given by
\begin{eqnarray}
\frac{\sqrt{n} k(0)}{e^{a/n}r_0}=k_{\rm GL} =  \frac{\sqrt{n}}{\rho_0} \left(1-\fr{2n}+\frac{7}{8n^2}\right).\label{eq:kgl-nnlo}
\end{eqnarray}
\begin{figure}[t]
\begin{center}
\includegraphics[width=7cm]{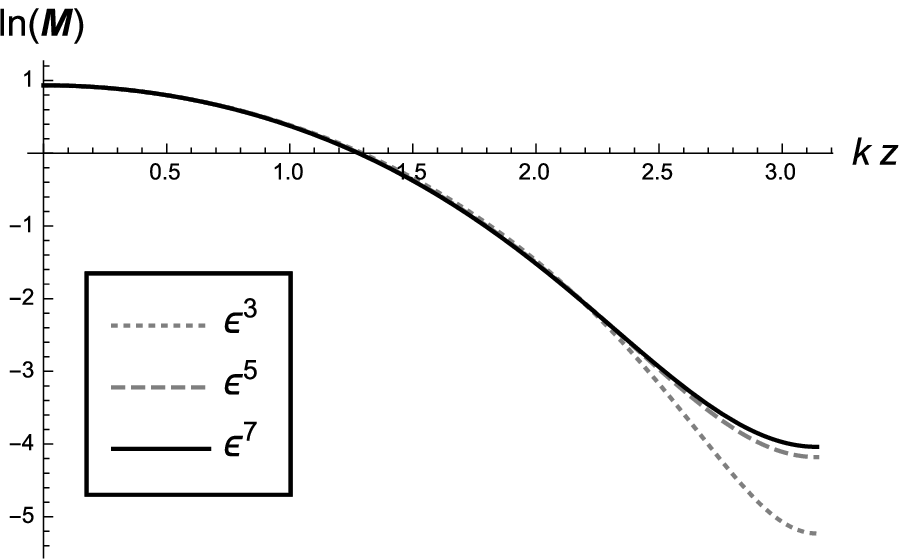}
\includegraphics[width=7cm]{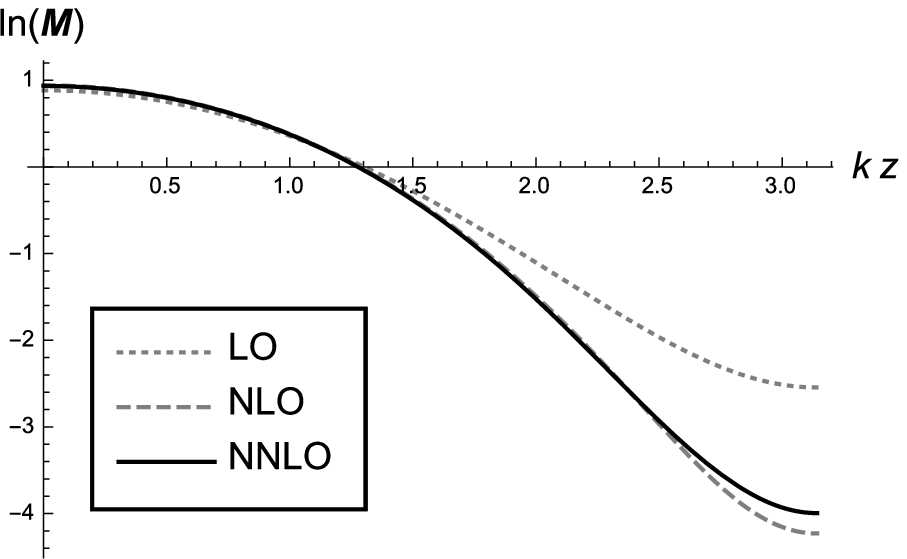}
\caption{The convergence of $\ln \tiM(z)$ for $n=10,\ \epsilon=1.2$ with respect to $\epsilon$ (left)
and $1/n$ (right). Only the half period is plotted. In the left, each curve corresponds to the NNLO
solution up to $\ord{\epsilon^3}$, $\ord{\epsilon^5}$ and  $\ord{\epsilon^7}$, respectively. In the right, each curve is the series solution up to $\ord{\epsilon^7}$ including the LO, NLO and NNLO correction. While the series in $\epsilon$ converges, the expansion in $1/n$ is not so good around the neck $\ln \tiM \sim -4$.}
\label{fig:conv}
\end{center}
\end{figure}

\begin{figure}[t]
\begin{center}
\includegraphics[width=8cm]{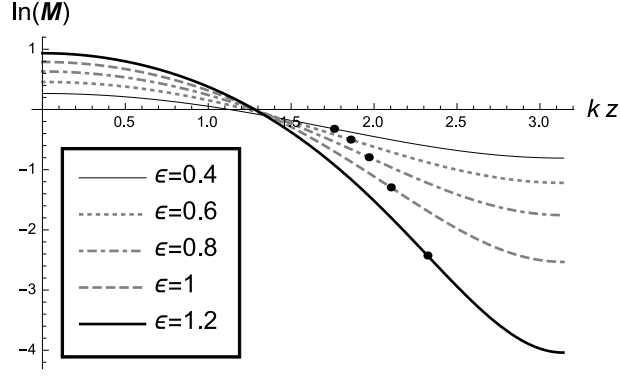}
\caption{The deformation for $n=10$ in a half period. The curves are for $\epsilon=0.4,0.6,0.8,1$ and $1.2$, each with the non-uniformity $\lambda=0.0566,0.0912,0.135,0.197$ and $0.316$, respectively.
The corresponding non-uniformities are calculated by eq.~(\ref{eq:non-uniform-param-def-inv}) with $\tib=1-\epsilon^2/2$.
Each dot on curves represents the point where the sign of $\partial_z^2 \ln \tiM(z)$ changes.}
\label{fig:deform1}
\end{center}
\end{figure}

The series solution in eq.~(\ref{eq:nnlo-ser-sol}) itself is convergent even around $\epsilon \simeq \sqrt{2}$ where the rescaled tension $\tib=1-\epsilon^2/2$ reaches zero, regardless of the convergence in the $1/n$ expansion (figure.~\ref{fig:conv}).
So, in the following analysis, we use this series solution, instead of solving eq.~(\ref{eq:horizon-eq-2nd}) numerically.
Since the parameter $a$ just involves the scaling, the shape of the deformation is governed only by $\epsilon$. Figure~\ref{fig:deform1} shows that as $\epsilon$ approaches to $\sqrt{2}$, the deformation tends to have the broader bulge and narrower neck as seen by the numerical calculations. Analytically, this can be shown by estimating the inflection point $\partial_z^2 \ln \tiM(z_0)=0$,
\begin{eqnarray}
&&k(\epsilon) z_0 - \frac{\pi}{2} = \epsilon\left(\fr{3}+\frac{1}{2n}(1+2\zeta(2)-4\ln 2)\right.\nonum
&& \quad\left.+\ \fr{2n^2}\left(\frac{258}{72}+3\zeta(2)+ 5\zeta(2)^2+10 (1-2\zeta(2))\ln2-12\ln^2 2+7\zeta(3)\right)\right)+\ord{\epsilon^3}\nonum
&&\qquad \simeq  \left(1+\frac{2.3}{n}+\frac{13.2}{n^2}\right)\frac{\epsilon}{3} + \ord{\epsilon^3}.
\end{eqnarray} 

\section{Thermodynamics}\label{sec:phase}
In this section, we study the thermodynamic properties of NUBS in the $1/n$ expansion.

\subsection{Variables}
First, we compute the thermodynamic variables from the solution~(\ref{eq:nnlo-ser-sol}).
The mass and tension are derived from eq.~(\ref{eq:ADM-mass-tension}).
Using eqs.~(\ref{eq:NNLOmatch-phiz}) and (\ref{eq:A0-nnlo}), the average of $\phi(z)$ can be
 expressed by the average of $\tiM(z)$
\begin{eqnarray}
&&\average{\phi} = \average{\cA_0} = 4e^a \left[\average{\tiM}+\frac{\average{\tiM}+\tib(\zeta(2)-1)}{n}\right.\nonum
&& \left.+\frac{\tib}{n^2}\left(\zeta(2)-\fr{2}\zeta(2)^2+2\zeta(3)-\left(1-2\zeta(2)+\frac{7}{10}\zeta(2)^2\right)\average{\left(\frac{\tiM'}{\tiM}\right)^2}\right)\right] \label{eq:phase-alpha-ave}.
\end{eqnarray}
where we eliminated $\ln \tiM(z)$ and higher derivatives of $\tiM(z)$ by using eq.~(\ref{eq:horizon-eq-2nd})
and integrating by part.
Using the series solution~(\ref{eq:nnlo-ser-sol}), the above averages can be computed as
\begin{eqnarray}
 \average{\tiM} =\mu_0(\epsilon),\quad \cout{\average{\left(\frac{\tiM'}{\tiM}\right)^2} = k(\epsilon)^2\frac{\epsilon^2 \mu_1^2(\epsilon)}{2\mu_0^2(\epsilon)} + \ord{\epsilon^4} \simeq \frac{\epsilon^2}{2},}
\average{\left(\frac{\tiM'}{\tiM}\right)^2} \simeq  \frac{\epsilon ^2}{2}+\frac{11 \epsilon ^4}{96}+\frac{127 \epsilon ^6}{3456}, \label{eq:M-averages}
\end{eqnarray}
For the latter, we also used $\mu_1(\epsilon),\mu_2(\epsilon), \mu_3(\epsilon)$ and eq.~(\ref{eq:nnlo-kgl}). The form of $\mu_0(\epsilon),\mu_1(\epsilon),\mu_2(\epsilon)$ and $\mu_3(\epsilon)$ is given in the appendix.~\ref{sec:nnloser}.
In the UBS limit $\epsilon\rightarrow 0$, we obtain the mass of UBS on the GL threshold point
\begin{eqnarray}
 {\cal M}_{\rm GL} = \underset{\epsilon \rightarrow 0}{\lim} \ {\cal M}  = \frac{(n+1)\omegaS_{n+1}L e^a r_0^n}{4\pi G}.
\end{eqnarray}
Comparing with eq.~(\ref{eq:ubs-mass-tension}), the length scale of the UBS mass on the threshold becomes
\begin{eqnarray}
 \rho_0 = (4e^{a})^{1/n}r_0.\label{eq:ubs-scale-rho}
\end{eqnarray}
On the other hand, since the horizon position $r_h(z) = r_0 M(z)^{1/n}$ varies in $z$, the horizon area is given by
\begin{eqnarray}
 {\cal A} =r_0^{n+1} \omegaS_{n+1}\int_0^L B_H(z)^{1/2}C_H(z)^\frac{n+1}{2}\sqrt{1+\frac{M(z)^{2/n}(M'(z))^2}{nM(z)^2}} M(z)^{1+1/n}dz
\end{eqnarray}
where $B_H(z)$ and $C_H(z)$ are the metric components on the horizon.
Using eq.~(\ref{eq:horizon-eq-2nd}) and integrating by part, the area just reduces to
\begin{eqnarray}
{\cal A} = 4^{1/n}e^{\frac{a}{n}}\omegaS_{n+1}r_0^{n+1} L \average{\phi}.
\end{eqnarray}
where we used the form of eq.~(\ref{eq:phase-alpha-ave}).
The surface gravity is also computed by using eq.~(\ref{eq:horizon-eq-2nd})
\begin{eqnarray}
\kappa = \frac{n}{2}\frac{1}{(4e^{a})^{1/n}r_0}. \label{eq:kappa-def}
\end{eqnarray}
The Smarr's formula just follows from eq.~(\ref{eq:ADM-mass-tension}) and the above expressions for ${\cal A}$ and $\kappa$,
\begin{eqnarray}
  \frac{(n+1){\mathcal M} -   {\mathcal T}L }{n+2} 
  = \frac{n\omegaS_{n+1}r_0^n L}{16\pi G}\average{\phi}= \frac{\kappa {\cal A}}{8\pi G}.\label{eq:smarr}
    \end{eqnarray}
In fact, the scaling factor can be factored out from these variables
\begin{eqnarray}
&& {\cal M} = \frac{\omegaS_{n+1}(4^{1/n} r_0)^{n+1}}{16\pi G}e^{\frac{n+1}{n}a} \frac{(n+2)n\phi_0(\epsilon)+1-\epsilon^2/2}{n+1}L_0(\epsilon),\nonum
&&   {\cal T} = \frac{\omegaS_{n+1}r_0^n}{4\pi G}e^a \left(1-\frac{\epsilon^2}{2}\right),\quad {\cal A} = \omegaS_{n+1}(4^{1/n}r_0)^{n+2} e^{\frac{n+2}{n}a}\phi_0(\epsilon) L_0(\epsilon)\label{eq:variables-sep},
\end{eqnarray}
where we introduced the scale invariant quantities
\begin{eqnarray}
 \phi_0(\epsilon) \equiv \frac{\average{\phi}}{4e^a},\quad L_0(\epsilon) \equiv \frac{L}{(4e^a)^{1/n}r_0} = \frac{2\pi}{4^{1/n}\sqrt{n}k(\epsilon)}.
\end{eqnarray}
Substituting eq.~(\ref{eq:M-averages}) into eq.~(\ref{eq:phase-alpha-ave}), $\phi_0(\epsilon)$ up to $\ord{\epsilon^6}$ is given by
 \begin{eqnarray}
\phi_0(\epsilon) = 1-\frac{\epsilon ^4}{96}-\frac{11 \epsilon ^6}{3456}+\fr{n}\left(\frac{\epsilon ^2}{2}+\frac{\epsilon ^4}{12}+\frac{79 \epsilon
   ^6}{3456}\right)+\fr{n^2}\left(\frac{\epsilon ^4}{48}+\frac{29 \epsilon ^6}{3456}\right).\label{eq:phi0}
\end{eqnarray}

\paragraph{First law}
Now, we check the first law for NUBS
\begin{eqnarray}
 d{\cal M} = \frac{\kappa}{8\pi G}d{\cal A}+{\cal T} dL.
\end{eqnarray}
The first law for the scaling $a$ is easily confirmed. For the variation with $\epsilon$, the first law reduces to the condition
\begin{eqnarray}
 \frac{d\phi_0(\epsilon)}{d\epsilon} - \frac{\epsilon}{n}+(\phi_0(\epsilon)-1+\epsilon^2/2) \frac{d\ln L_0(\epsilon)}{d\epsilon} = 0.\label{eq:1stlaw}
\end{eqnarray}
From eqs.~(\ref{eq:phi0}) and (\ref{eq:nnlo-kgl}), this condition can be confirmed at least for up to $\ord{\epsilon^6}$ and $\ord{n^{-2}}$.

\subsection{Circumference}
So far, we have considered the scaling and deformation separately.
However, we have to vary the scaling $a$ together with $\epsilon$
to obtain the physical phase diagram.
If we recover the scaling, the circumference of the extra dimension is given by
\begin{eqnarray}
 L = \frac{2\pi e^{a/n}  r_0}{\sqrt{n} k(\epsilon)}.
\end{eqnarray}
Therefore, the circumference also changes as the solution gets deformed.
To obtain the sequence of NUBS with a fixed $L$, the scaling $a$ should be tuned so that the running of $\epsilon$ is canceled. Since the running in $k(\epsilon)$ becomes comparable to $e^{a/n}$
only if $\epsilon^2 \sim n^{-1}$, we introduce a rescaled non-uniformity parameter
\begin{eqnarray}
   \epsilon^2 = \frac{2s}{n},
\end{eqnarray}
where we assume $s\sim\ord{1}$.
With this assumption, the variation in eq.~(\ref{eq:nnlo-kgl}) is cancelled by setting
\begin{eqnarray}
 a = -\frac{s}{12}+ \fr{n}\left(\frac{3s}{4}-\frac{23s^2}{576}\right)-\fr{n^2}\left(\frac{7s}{12}-\frac{11s^2}{32}+\frac{997s^3}{38880}\right),\label{eq:s-Lfix}
 \end{eqnarray}
 where the scaling at $s=0$ is absorbed in $r_0$.
 Therefore,  the physical phase of NUBS is expressed in terms of an $\ord{1}$ non-uniformity parameter $s$.

\subsection{Phase diagram}
Now, we investigate the phase diagram for the constant $L$.
 With the assumption $\epsilon^2 = 2s/n$ and the scaling in eq.~(\ref{eq:s-Lfix}), $L$ is now fixed to
 \begin{eqnarray}
  L=L_{\rm GL} \equiv \frac{2\pi}{k_{\rm GL}}
\end{eqnarray}
where $k_{\rm GL}$ is given by eq.~(\ref{eq:kgl-nnlo}).
From eqs.~(\ref{eq:phi0}) and (\ref{eq:s-Lfix}),
 eqs.~(\ref{eq:variables-sep}) and (\ref{eq:kappa-def}) gives
\begin{subequations}\label{eq:thermo-result}
\begin{align}
 \frac{{\cal M}}{{\cal M}_{\rm GL}} &= e^{-s/12}\left(1+\left(1+\frac{5}{9n}\right)\frac{3s}{4n}-\left(\frac{23}{576}-\frac{7}{12n}\right)\frac{s^2}{n}-\frac{17291s^3}{311040n^2}+\frac{529s^4}{663552n^2}\right),
 \label{eq:thermo-result-M}\\
\frac{{\cal T}}{ {\cal T}_{\rm GL}} &= e^{-s/12}\left(1-\left(1+\frac{7}{3 n}\right) \frac{s}{4n}
-\left(\frac{23}{576 }+\frac{1}{8 n}\right) \frac{s^2}{n}-\frac{4871 s^3}{311040 n^2}
+\frac{529 s^4}{663552 n^2}\right)\label{eq:thermo-result-T},\\
 \frac{{\mathcal A}}{{\mathcal A}_{GL}} &= e^{-s/12}\left(1+\left(\frac{2}{3}+\frac{7}{6n}\right)\frac{s}{n}
 -\left(\frac{23}{576}-\frac{31}{64n}\right)\frac{s^2}{n}-\frac{127s^3}{2430n^2}+\frac{529s^4}{663552n^2}\right)\label{eq:thermo-result-A},\\
 \frac{\kappa}{\kappa_{GL}}&=1+\left(1-\frac{9}{n}+\frac{7}{n^2}\right) \frac{s}{12n}
 +\left(\frac{25}{576}-\frac{13}{32 n}\right)
   \frac{s^2}{n^2} +\frac{9041s^3}{311040 n^3}\label{eq:thermo-result-k},
\end{align}
\end{subequations}
where the value at the GL point is given by
\begin{eqnarray}
\begin{split}
{\cal M}_{\rm GL} &\equiv \frac{(n+1)\omegaS_{n+1}\rho_0^n L_{\rm GL}}{16\pi G} ,\quad {\cal T}_{GL} \equiv \frac{{\cal M}_{\rm GL}}{(n+1)L_{\rm GL}},\\
 {\cal A}_{\rm GL} &\equiv \omegaS_{n+1}\rho_0^{n+1}L_{\rm GL},\quad \kappa_{\rm GL} \equiv \frac{n}{2\rho_0}.
 \end{split}
\end{eqnarray}
\begin{figure}[h]
\begin{center}
\includegraphics[width=7.5cm]{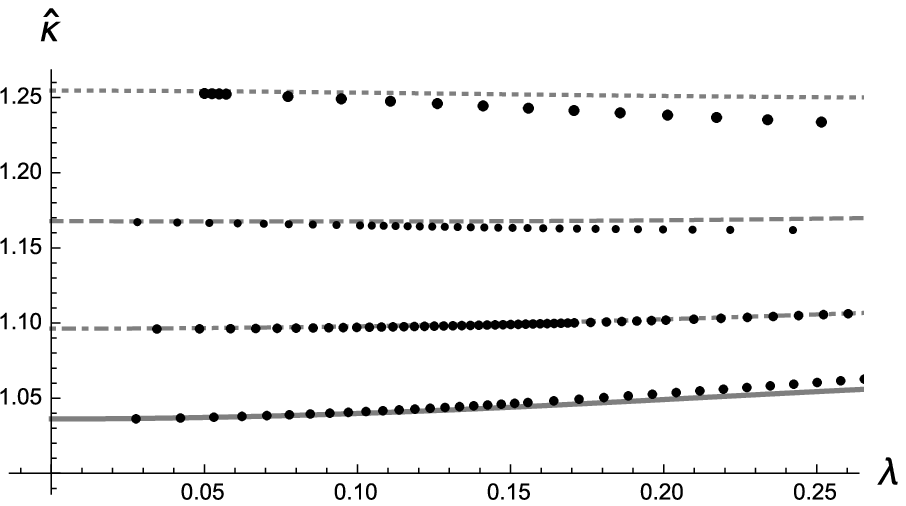}
\includegraphics[width=7.5cm]{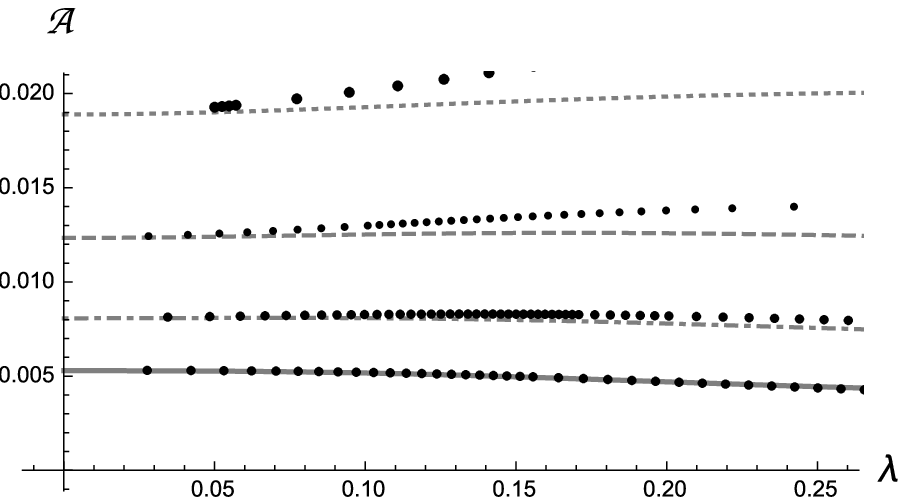}
\end{center}
\begin{center}
\includegraphics[width=7.5cm]{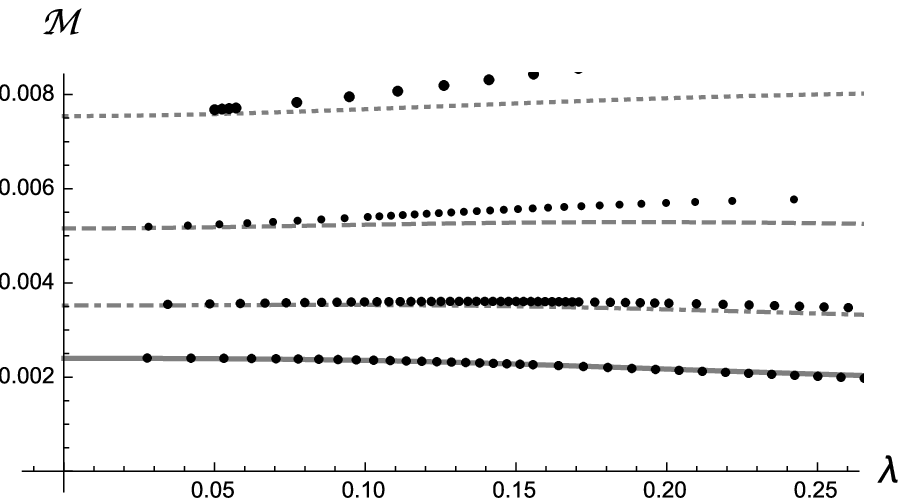}
\includegraphics[width=7.5cm]{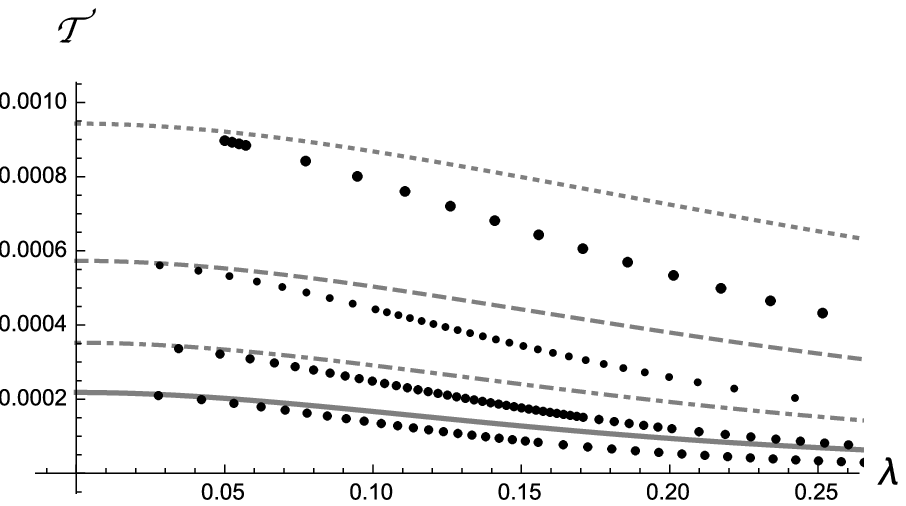}
\caption{Phase diagram of NUBS, and comparison with the numerical construction of~\cite{FiguerasMR12} (the data for the tension, not included in~\cite{FiguerasMR12}, has been provided by Pau Figueras). The normalized surface gravity $\hat{\kappa}=\kappa/n$ (top left), horizon area ${\cal A}$ (top right), mass ${\cal M}$ (bottom left)
 and tension ${\cal T}$ (bottom right)
 are plotted against the non-uniformity parameter $\lambda$. Each quantity is normalized in the unit of the extra dimension size $L$. The dotted, dashed, dot-dashed and solid curves are the variables for $n=7,8,9,10\ (D=11,12,13,14)$, respectively.}\label{fig:compare}
\end{center}
\end{figure}

In figure~\ref{fig:compare}, we compare our result with and the numerical result in~\cite{FiguerasMR12} at several dimensions.
The corresponding non-uniformity parameter $\lambda$ is determined by eq.~(\ref{eq:non-uniform-param-def-inv}) with $\tib=1-s/n$.
The variables are plotted in the unit of $L_{\rm GL}$,
\begin{eqnarray}
\begin{split}
 \frac{{\cal M}}{(L_{\rm GL})^{n+1}}& = \frac{(n+1)\omegaS_{n+1}}{16\pi}\left(\frac{\bar{k}_{\rm GL}}{2\pi}\right)^n\left( \frac{{\cal M}}{{\cal M}_{\rm GL}} \right),\quad   \frac{{\cal T}}{(L_{\rm GL})^{n}} = \frac{\omegaS_{n+1}}{16\pi}\left(\frac{\bar{k}_{\rm GL}}{2\pi}\right)^n \left(\frac{{\cal T}}{{\cal T}_{\rm GL}}\right) 
\\
  \frac{{\cal A}}{(L_{\rm GL})^{n+2}} &= \omegaS_{n+1}\left(\frac{\bar{k}_{\rm GL}}{2\pi}\right)^{n+1}\left( \frac{{\cal A}}{{\cal A}_{\rm GL}} \right) ,\quad \kappa L_{\rm GL} = \frac{n\pi}{\bar{k}_{\rm GL}}\left(\frac{\kappa}{\kappa_{\rm GL}}\right)
\end{split}
\end{eqnarray}
where we write $\bar{k}_{\rm GL} = \rho_0 k_{\rm GL}$ and set $G=1$.
The formula in eq.~(\ref{eq:thermo-result}) reproduces the numerical result within the error of $\ord{n^{-3}}$ for the small non-uniformity. For the large non-uniformity, the $1/n$ expansion breaks down and one can  no longer expect the good accuracy.

In figure~\ref{fig:compare}, the tension seems to have the relatively large error compared to other variables.
Actually, the Smarr formula~(\ref{eq:smarr}) implies that the tension has the error of one lower order 
than other variables. This is because the tension is sensitive to the inhomogeneity of the horizon, especially the neck part of the horizon which has relatively bad convergence (figure~\ref{fig:conv}). On the other hand, the mass and area,  which are the averaged quantities, will not be so sensitive to the neck geometry. The surface gravity, which is defined uniformly on the horizon, also will not have such sensitivity.

\subsubsection{Localized black holes}\label{sec:smallbh}
We should note that our formalism also includes localized black holes as the zero tension solution.
If the solution has the point $M(z)=0$, the regularity in eq.~(\ref{eq:LO-sol-B}) requires
\begin{eqnarray}
\left( M'(z)^2 - \frac{M''(z)}{M(z)}\right)_{M(z)=0} = 0.
\end{eqnarray}
From the effective equation~(\ref{eq:LO-eff-const}), this is the case of the zero tension $\beta=0$.
For $\beta=0$, eq.~(\ref{eq:horizon-eq-2nd}) admits an analytic solution
\begin{eqnarray}
 M(z) = e^{-z^2/2}\left(1-\frac{z^4}{n}+\frac{z^6(-16+3z^2)}{96n^2}\right)
\label{eq:sbh-sol}
\end{eqnarray}
where we omit the scaling for simplicity.
This solution is not periodic nor convergent at the large $z$, so one can think this unphysical.
But, this can be interpreted as the near-equatorial part of the spherical solution.

Let us consider $D=n+4$ Schwarzshild spacetimes in the isotropic coordinate
\begin{eqnarray}
 ds^2 = -\left(\frac{\rho^{n+1}-\rho_0^{n+1}}{\rho^{n+1}+\rho_0^{n+1}}\right)^2 dt^2 + \left(1+\left(\frac{\rho_0}{\rho}\right)^{n+1}\right)^\frac{4}{n+1}(d\rho^2+\rho^2d\theta^2+\rho^2 \sin^2\theta d\Omega_{n+1}^2).\nonum \label{eq:sbh-isotropic}
\end{eqnarray}
We consider the transformation to the cylindrical coordinate
\begin{eqnarray}
 r= \rho \sin\theta,\quad r_0 z = \sqrt{n} \rho \cos\theta,
\end{eqnarray}
where $z$ coordinate describes the region around the equatorial plane $\theta \approx \pi/2$.
Then, we obtain
\begin{eqnarray}
 ds^2 \simeq -\left(\frac{\sR-e^{-z^2/2}}{\sR+e^{-z^2/2}}\right)^2 dt^2 + \left(1+\frac{e^{-z^2/2}}{\sR}\right)^\frac{4}{n+1}\left(dr^2+\frac{r_0^2dz^2}{n}\right)+r^2  \left(1+\frac{e^{-z^2/2}}{\sR}\right)^\frac{4}{n+1}d\Omega_{n+1}^2\nonum
\end{eqnarray}
where we set $\rho_0 = r_0$ and the following formula is used
\begin{eqnarray}
 \sin^{n+1}\theta = \left(1+\frac{z^2}{n\sR^{2/n}}\right)^{-\frac{n+1}{2}} \simeq e^{-\frac{z^2}{2}} + \ord{n^{-1}}.\label{eq:sbh-sin-match}
\end{eqnarray}
This is identical to what we have obtained in eqs.~(\ref{eq:LO-sol-AC}) and (\ref{eq:LO-sol-B}) with the leading order of eq.~(\ref{eq:sbh-sol}).

As for the higher order, though it is difficult to find the coordinate transformation,
Eq.~(\ref{eq:sbh-sol}) reproduces the mass spectrum of localized black holes, correctly up to NNLO.\footnote{We thank Roberto Emparan for suggesting this possibility.}
Since we have $\beta=0$, substituting eq.~(\ref{eq:phase-alpha-ave}) into eq.~(\ref{eq:ADM-mass-tension}) gives
\begin{eqnarray}
 {\cal M} &=& \frac{(n+2)\Omega_{n+1}\rho_0^n}{4\pi G} L\average{M} \nonum
 &=& \frac{(n+2)\Omega_{n+1}\rho_0^{n+1}}{4\pi G} \int_{-\infty}^{\infty} M(z) \frac{dz}{\sqrt{n}}\nonum
 &=& \frac{(n+2)\Omega_{n+1}\rho_0^{n+1}}{4\pi G} \sqrt{\frac{2\pi}{n}}\left(1-\frac{3}{4n}+\frac{25}{32n^2}\right)\label{eq:mass-sbh}
\end{eqnarray}
where we identified $L\average{M}$ with the integration over $Z=r_0 z/\sqrt{n}$ in the middle line.
Using the following identity
\begin{eqnarray}
 \frac{\Omega_{n+2}}{\Omega_{n+1}} \simeq \sqrt{\frac{2\pi}{n}}\left(1-\frac{3}{4n}+\frac{25}{32n^2}\right),
\end{eqnarray}
eq.~(\ref{eq:mass-sbh}) reproduces the mass of $D=n+4$ Schwarzschild black holes in the isotropic coordinate~(\ref{eq:sbh-isotropic}) up to NNLO, despite the solution~(\ref{eq:sbh-sol}) itself breaks down at the large $z$.
As discussed in~\cite{EmparanSSTT15}, this is because the contribution from the equatorial region
becomes dominant for the energy density in the large $D$ limit.

However, our construction fails to capture the thermodynamics and deformation in the localized black hole phase. Eq.~(\ref{eq:mass-sbh}) is only the mass of the spherical black hole without the deformation.
If we compere the mass with the NUBS and UBS phase for the same $L$, any size of the localized black hole
has the zero mass, because of the infinite period. Then, in the large $D$ limit, the localized black hole phase is degenerate
to a single point of the phase diagram: (${\cal M}=0, {\cal T}=0$).
This is because the deformation effect is non-perturbative in $1/n$ for localized black holes,
at least, in the small gradient assumption.
The deformation of the localized black hole with the radius $\rho_0$ and extra dimension size $L$ is supported by the gravitational potential from itself at a distance of a  half period $L/2$.
This gravitational effect has the magnitude of $(2\rho_0/L)^{D}$, which is negligible in the $1/D$ expansion, unless the black hole is very large $L/2-\rho_0\sim \rho_0/D$.

\subsection{Critical dimension}
\begin{figure}[h]
\begin{center}
\includegraphics{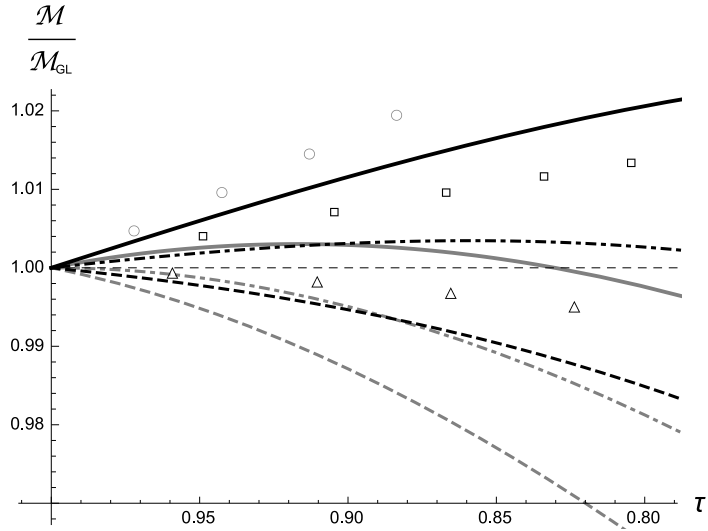}
\caption{Critical behavior of NUBS near the bifurcation point, in which the normalized mass is plotted against the relative tension. The thick, dot-dashed and dashed curves are for $n=8,9,10$ $(D=12,13,14)$, respectively.  We plotted the results up to the NLO (gray) and NNLO (black) corrections,
in comparison with the numerical data for $n=8$ (circle), $n=9$ (square) and $n=10$ (triangle) in~\cite{FiguerasMR12} (the data for the relative tension, not included in~\cite{FiguerasMR12}, is produced with the tension data provided by Pau Figueras). At NNLO, the mass seems to grow by the effect of the self-interaction, which makes the $n=9$ phase bellow the critical dimension.}
\label{fig:nubs-phase}
\end{center}
\end{figure}

The phase diagram of NUBS admits a critical behavior between $D=13$ and $D=14$,
across which the order of the transition changes~\cite{Sorkin04}.
This means that NUBS has the greater mass than UBS on the GL point for $D\leq13$,
and the lower mass for $D>13$.
In eq.~(\ref{eq:thermo-result-M}), the gradient near the bifurcation point becomes
\begin{eqnarray}
 \frac{{\cal M}}{{\cal M}_{\rm GL}} \simeq 1- \left(1-\frac{9}{n}-\frac{5}{n^2}\right)\frac{s}{12},\label{eq:criticalMs}
\end{eqnarray}
in which the gradient changes the sign at 
\begin{eqnarray}
 n^2 -9n-5 = 0\label{eq:criticalcond1}.
\end{eqnarray}
This implies the existence of the critical dimension at $n_*\simeq 9.5$ ($D_*\simeq13.5$), which agrees with the Sorkin's numerical result~\cite{Sorkin04}, within the error of $\ord{n^{-2}}$. Here we note that the $n=9$ case
admits a marginal behavior at NLO, in which case the NNLO correction plays a crucial role.\footnote{A subtlety is that the leading and sub-leading terms become comparable around the critical dimension. One may suspect the validity of the expansion. But we believe this only happens at the leading order, since the subsequent correction seems convergent.}
In figure~\ref{fig:nubs-phase}, we show the critical behavior in the phase diagram,
plotted in terms of the normalized mass and relative tension.
The relative tension is a dimensionless quantity defined by
\begin{eqnarray}
\tau = \frac{(n+1)L{\cal T}}{{\cal M}} = \frac{(n+1)^2(1-\epsilon^2/2)}{(n+2)n\phi_0(\epsilon)+1-\epsilon^2/2}\label{eq:tau-ep}
\end{eqnarray}
which is normalized so that the UBS limit gives $\tau = 1$.
Substituting $\epsilon^2 = 2s/n$, we obtain
\begin{eqnarray}
\tau = 1-\frac{s}{n}\left(1+\fr{n}-\fr{n^2}\right)+\frac{s^2}{24n^2}\left(1+\frac{16}{n}\right)-\frac{7s^3}{432n^3},
\label{eq:tau-s-def}
\end{eqnarray}
where the corrections can be determined up to $\ord{n^{-3}}$,
 because $\ord{n^{-3}}$ correction in eq.~(\ref{eq:tau-ep}) should start with $\epsilon^2$ for the normalization $\tau(\epsilon=0)=1$.
Substituting this expression into eq.~(\ref{eq:criticalMs}), we obtain the gradient in the phase diagram,
\begin{eqnarray}
 \frac{{\cal M}}{{\cal M}_{\rm GL}} \simeq  1-\frac{n}{12} \left(1-\frac{10}{n}+\frac{6}{n^2}\right)(1-\tau).
\end{eqnarray}
The condition $n^2-10n+6=0$ also gives $n_*\simeq9.4$. Since the gradient in eq.~(\ref{eq:tau-s-def}) does not affect the sign, this value is identical to $n_*\simeq 9.5$
 up to the error in $\ord{n^{-2}}$.

The horizon area is also known to admit the critical behavior at the same dimension as the mass.
Above the critical dimension, the horizon area of NUBS becomes greater than that of UBS with the same mass, which means that NUBS is more favored than UBS in the microcanonical ensemble. Below the critical dimension, NUBS becomes the unstable phase. 
However, we cannot observe the difference in the area between NUBS and UBS of the same mass to this order,
\begin{eqnarray}
 {\mathcal A}/{\mathcal A}_{\rm UBS} = 1+ \ord{n^{-3}}
\end{eqnarray}
where the horizon area of the UBS with the same mass is given by
\begin{eqnarray}
 {\cal A}_{\rm UBS} = \omegaS_{n+1} L_{\rm GL} \rho_0^{n+1}e^{\frac{n+1}{n}a} \left[\frac{(n+2)n\phi_0(\epsilon)+1-\epsilon^2/2}{(n+1)^2}\right]^\frac{n+1}{n}.
\end{eqnarray}
This is expected since the first law guarantees the cancelation of the linear order dependence~\cite{Gubser02,Sorkin04}. In this case, both ${\cal A}$ and ${\cal A}_{\rm UBS}$ are the function of $\epsilon^2$, and then the second order becomes $\epsilon^4$.
In addition, the coefficient will be at most $\ord{n^{-1}}$, since there is no entropy difference in the large $D$ limit~\cite{EmparanST13}.
Then, the leading behavior of the area difference should appear at much higher order than the mass
\begin{eqnarray}
  {\mathcal A}/{\mathcal A}_{\rm UBS}-1 \sim \ord{\epsilon^4/n} \sim  \ord{n^{-3}}.
\end{eqnarray}
The numerical calculation actually shows
${\cal A}/{\cal A}_{\rm UBS}-1 \simeq 10^{-3}\sim10^{-4}$ for the small non-uniformity in $D=11$ to $14$ (See figure 7 in~\cite{FiguerasMR12}).

As for the surface gravity in eq.~(\ref{eq:thermo-result-k}), though its thermodynamical interpretation is unclear, 
we see that the gradient at the GL point changes at
\begin{eqnarray}
 n^2-9n+7=0,
\end{eqnarray}
which gives a smaller critical value $n_*\simeq 8.1$ ($D_*\simeq12.1$). Then, the NUBS phase becomes hotter in $D>12$ and cooler in $D\leq 12$. This effect is also observed by the numerical calculation~\cite{Sorkin04,FiguerasMR12}.

\section{Summary}\label{sec:sum}
In the large $D$ limit,
the near-horizon dynamics with the small gradient ($\ll D/r_0$) along the horizon
reduces to the nonlinear effective theory for some collective degrees of freedom of the horizon~\cite{EmparanSSTT15,Minwalla+15}. This effective approach will be a powerful tool for the various nonlinear problems in the higher dimensional gravity.
 
In this paper, we applied the large $D$ effective theory approach
to the study of the non-uniform black strings (NUBS)
bifurcating from the uniform black strings (UBS) on the threshold of the Gregory-Laflamme (GL) instability.
 Extending the leading order analysis~\cite{EmparanSSTT15}, we solved the near-horizon
geometry up to the next-to-next-to-leading order (NNLO) in the $1/D$ expansion.
We also analyzed the far-zone geometry by the Newtonian approximation
and derived the matching condition between the two zones.
 As the result, we obtained the effective equation of the deformed horizon up to NNLO.
We also obtained the thermodynamic variables up to NNLO from the matching result.
The phase diagram produced by the large $D$ analysis agrees with the numerical results for the small non-uniformity within the expected precision.
We also show that our construction contains localized black holes
which have the degenerate phase in the $1/D$ expansion.
The phase diagram admits the critical behavior in the mass around $D_*\simeq13.5$ as shown numerically~\cite{Sorkin04,FiguerasMR12}.
We cannot find the area difference between NUBS and UBS of the same mass in this order.
It will be straightforward (though much harder) to obtain much higher order corrections,
 in which the critical behavior in the area difference will be observed.
 
 In the higher dimension, various black objects admit the GL-like instabilities and
their zero modes are expected to produce the new deformed branches.
For example, Myers Perry black holes were conjectured to be unstable for the sufficiently large rotation~\cite{Emparan:2003sy}, and it is numerically shown that each zero mode deformation leads to
  the new `bumpy' solution~\cite{Dias:2014cia,Emparan:2014pra}.
AdS$_5\times S^5$ black holes also have the deformed blanches~\cite{Dias:2015pda}.
These deformed solutions will show the similar critical behavior.
The large $D$ effective theory approach will be a good analytical tool to observe such nonlinear phenomenon.
The authors have already shown that the bumpy Myers-Perry solution can be obtained by the $1/D$ expansion~\cite{SuzukiTanabe15a}.

As the further extension of this work, it is interesting to study the time evolution of the non-uniformity in the $1/D$ expansion.
This will be partially realized by including the slow time dependence, which is already formulated in \cite{Minwalla+15}. For $D>13$, the NUBS phase becomes stable and the time evolution of the GL instability can be followed by the current small gradient effective approach. Since the numerical simulation becomes more difficult in the higher dimension by the resolution problem, the large $D$ analysis will be a suitable approach to this problem.
For $D\leq13$, the NUBS phase is no longer stable, and the GL instability develops to the self-similar structure which consists of the sequence of black holes and thin strings connected with each others~\cite{Choptuik03,Garfinkle04,LehnerPretorius10}. Since the small gradient assumption will break down in such phase, we need to incorporate the large gradient ($\sim D/r_0$) dynamics which is not yet
formulated. 
 
The large gradient dynamics will be also required to understand static solutions supported by the interaction between the multiple horizons, because the small gradient degrees of freedom are confined in the near-horizon geometry.
In section~\ref{sec:smallbh}, we show that  localized black holes have the degenerate phase
in the large $D$ limit. This is because the gravitational force becomes non-perturvative in $1/D$, $(2\rho_0/L)^D$.
To keep the deformation finite in the large $D$ limit, one should assume very large black holes $L/2-\rho_0 \sim  \rho_0/D$, which will involve the large gradient dynamics.
Near the merger point, the NUBS phase was conjectured to have the conical waist~\cite{Kol:2002xz,Emparan:2011ve}. In such phase, the gradient along the horizon will be also large.

\section*{Acknowledgement}
The authors are very grateful to Roberto Emparan for valuable comments and discussions, and Pau Figueras, Keiju Murata and Harvey S. Reall for the kind supply of the numerical data used for the comparison. KT was supported by JSPS Grant-in-Aid for Scientific Research No.26-3387.

\appendix
\section{Uniform Black Strings in the conformal coordinate}
In this section, we will show the large $D$ limit of the uniform black string in the conformal coordinate.
$D=n+4$ black string solutions are given by
\begin{eqnarray}
	ds^2=-f(\rho)dt^2+f(\rho)^{-1}dr^2+\rho^2 d\Omega_{n+1}^2+dZ^2,
	\quad f(\rho)=1-\frac{\rho_0^{n}}{\rho^{n}}. \label{eq:ubs-ordinary}
\end{eqnarray}
If the extra dimension is compactified with the circumference $L$, the mass and tension of the spacetime are given by
\begin{eqnarray}
 {\cal M} = \frac{(n+1)\omegaS_{n+1}L\rho_0^n}{16\pi G}, \quad {\cal T} = \frac{\omegaS_{n+1}\rho_0^n}{16\pi G}\label{eq:ubs-mass-tension}
\end{eqnarray}
where $\omegaS_{n+1}=2\pi^{n/2+1}/\Gamma(n/2+1)$ is the volume of $S^{n+1}$.
This metric can be transformed to the conformal coordinate
\begin{eqnarray}
		ds^2 = - F^2(r)dt^2 + dr^2+dZ^2+\rho^2(r) d\Omega_{n+1}^2\label{eq:ubs-conformal0}.
\end{eqnarray}
by the transformation
\begin{eqnarray}
	\rho \rightarrow r =  \rho\ {}_2 F_1\left[\fr{2},-\fr{n},1-\fr{n};\left(\frac{\rho_0}{\rho}\right)^{n}\right],
\end{eqnarray}
where $r$ coincides with $\rho$ at $\rho \rightarrow \infty$.
$F(r)$ is implicitly given by
\begin{eqnarray}
 r =\frac{\rho_0}{(1-F(r)^2)^\fr{n}}\ {}_2 F_1\left[\fr{2},-\fr{n},1-\fr{n};1-F(r)^2\right].\label{eq:ubs-F-def}
\end{eqnarray}
\cout{
For the orthogonality of $r$ and $z$, $\partial_\rho \partial_z r =0$ should hold, which means $r = \chi(\rho)+\xi(z)$.
Substituting $r=\chi(\rho)+\xi(z)$ into eq.~(\ref{eq:ubs-ordinary}) and identifying with eq.~(\ref{eq:ubs-conformal0}), the following conditions are obtained
\begin{eqnarray}
	 H(r(\rho,z),z) \chi'(\rho) =f(\rho)^{-1/2} ,\quad H(r(\rho,z),z) (\xi'(z)^2 + 1) =1.
\end{eqnarray}
These conditions require $H$ to be a constant,
\begin{eqnarray}
	\chi'(\rho)f(\rho)^{1/2}=\xi'(z)^2+1=H^{-1}={\rm const.}
\end{eqnarray}
Then, also imposing $r \rightarrow \rho$ at $\rho \rightarrow \infty$ and the isometry with $z$, we obtain
	\begin{eqnarray}
	r=\chi(\rho) =  \rho\ {}_2 F_1\left[\fr{2},-\fr{n},1-\fr{n};\left(\frac{\rho_0}{\rho}\right)^{n}\right],\quad \xi(z)=0,\quad H(r,z)=1.
	\end{eqnarray}
$F(r)$ is given implicitly
\begin{eqnarray}
 r =\frac{\rho_0}{(1-F(r)^2)^\fr{n}}\ {}_2 F_1\left[\fr{2},-\fr{n},1-\fr{n};1-F(r)^2\right]\label{eq:ubs-F-def}
\end{eqnarray}
}
In this coordinate, the horizon position $F(r_h)=0$ becomes
\begin{eqnarray}
r_h =\frac{\rho_0\pi^{1/2}\Gamma\left(1-\fr{n}\right)}{\Gamma\left(\fr{2}-\fr{n}\right)}.\label{eq:ubs-r0}
\end{eqnarray}

\paragraph{Large D limit}
Though eq.~(\ref{eq:ubs-F-def}) cannot be solved explicitly, the large $n$ limit allows
the expression in the $1/n$ expansion.
Keeping $\sR=4r^n/\rho_0^n$ finite, the limit $n\rightarrow \infty$ in eq.~(\ref{eq:ubs-F-def})
leads to
\begin{eqnarray}
	\sR^{1/n}&=&\frac{4^{1/n}}{(1-F^2)^\fr{n}}\ {}_2 F_1\left[\fr{2},-\fr{n},1-\fr{n};1-F^2\right]\nonum
	&=&1-\frac{2}{n}\tanh^{-1} F+\ord{n^{-2}}\label{eq:ubs-F-largen}
\end{eqnarray}
Comparing $\ord{n^{-1}}$ terms in eq.~(\ref{eq:ubs-F-largen}), the leading term in $F(R)$ is determined
\begin{eqnarray}
F(\sR) = \frac{\sR-1}{\sR+1}+\ord{n^{-1}}.
\end{eqnarray}
Recalling the relation $F^2(\sR)=1-\rho_0^n/\rho^{n}$, we also obtain
\begin{eqnarray}
 \rho(\sR) = \rho_0 \left(1-F^2(\sR)\right)^{-\frac{1}{n}}=r_0 \sR^{1/n}\left(1+\fr{\sR}\right)^{2/n}+\ord{n^{-2}},
\end{eqnarray}
where we introduced the horizon scale $r_0 = 4^{-1/n}\rho_0$.
Thus, in the conformal coordinate, the near-horizon geometry of the uniform black string in the leading order becomes
\begin{eqnarray}
 ds^2 \simeq -\left(\frac{\sR-1}{\sR+1}\right)^2dt^2+dr^2+dZ^2+ r_0^2 \sR^{2/n} \left(1+\fr{\sR}\right)^{4/n}d\Omega_{n+1}^2.\label{eq:ubs-sol-LO}
\end{eqnarray}
We note that the horizon radius $r_h$ in eq.~(\ref{eq:ubs-r0}) slightly differs to $r_0$ in the higher order,
\begin{eqnarray}
 r_h \simeq r_0 \left(1-\frac{\zeta(2)}{n^2}-\frac{2\zeta(3)}{n^3}\right).\label{eq:ubs-rh-r0}
\end{eqnarray}

\section{Near-zone solution}\label{sec:NLO-sol}
\subsection{Next-to-leading order}
In the following, we use $\sX=\sR/M(z)$ for the abbreviation.
$\Li_k(x)$ denotes the polylogarithm.
\begin{eqnarray}
 &&A^{(1)} = - \frac{4M''(z)\sX(\sX-1)\ln \sX }{M(z)(\sX+1)^3} \nonum
 &&+\ \frac{4(M(z)^2-M'(z)^2+M(z)M''(z))\sX(\sX-1)(\zeta(2)+\ln^2\sX+2\ \Li_2(-\sX))}{M(z)^2(\sX+1)^3},
\end{eqnarray}
\begin{eqnarray}
 &&C^{(1)} = 8\ln^2(\sX+1)-\frac{4(M'(z)^2-M(z)M''(z))}{M(z)^2}\ln(\sX+1)\nonum
 && +\ ((3+5\sX)M(z)^2-(\sX-1)M'(z)^2+(\sX-1)M(z)M''(z))\frac{2\ln^2\sX}{(\sX+1)M(z)^2}\nonum
 &&-\ \ln \sX \left(16\ln (\sX+1)-\frac{4((\sX+1)M'(z)^2-\sX M(z)M''(z))}{(\sX+1)M(z)}\right)\nonum
 &&+\  \frac{4(M(z)^2-M'(z)^2+M(z)M''(z))}{(\sX+1)M(z)^2}\left(\zeta(2)\sX+(\sX-1)\Li_2(-\sX)\right)
 ,
\end{eqnarray}
and
\begin{eqnarray}
 &&B^{(1)} =\frac{8(M'(z)^2-M(z)M''(z))^2}{M(z)^4}\ln^2(\sX+1)\nonum
 &&+\ \frac{2(M(z)M''(z)-2M'(z)^2)(M(z)^2-M'(z)^2+M(z)M''(z))}{3M(z)^4}(\ln^3 \sX+6\ \Li_3(-\sX))\nonum
 &&+\  \frac{4(M(z)M''(z)-M'(z)^2)(M(z)^2-M'(z)^2+M(z)M''(z))}{M(z)^4}\nonum
 &&\qquad \times \left(2\ \Li_2\left(\frac{1-\sX}{2}\right)-\Li_2(1-\sX)+\ln^2 2+\frac{\zeta(2)\sX-2\sX\  \Li_2(-\sX)}{\sX+1}\right)\nonum
 &&-\ \frac{2 \ln ^2\sX}{(\sX+1) M(z)^4} \Bigl((\sX-1) M(z)^3 M''(z)+(-5 \sX-7) M'(z)^4\nonum
 &&\qquad +\ 4 (2 \sX+3) M(z) M'(z)^2
   M''(z)+M(z)^2 \left((\sX+3) M'(z)^2-(3 \sX+5) M''(z)^2\right)\Bigr)\nonum
   &&+\ \frac{4 \ln (\sX+1)}{M(z)^4} \Bigl(M(z)^3 M''(z) \left(-2 \ln 2 -2 \ln M(z)+\zeta(2)\right)+2
   \left(\zeta(2)-\ln 2 -1 \right) M'(z)^4\nonum
   &&\qquad-\ \left(-4 \ln2 +3\zeta(2)-3\right) M(z)
   M'(z)^2 M''(z)+M(z)^2 \left(\left(-2 \ln2 +\zeta(2)-1\right) M''(z)^2\right.\nonum
   &&\qquad \left.-\  M'(z)^2   \left(-2\ln2 -2\ln M(z)+2\zeta(2)+1\right)\right)\Bigr)\nonum
   &&+\ \frac{4 \ln \sX \ln (\sX+1) \left(-5 M'(z)^4+9 M(z) M'(z)^2 M''(z)+M(z)^2 \left(M'(z)^2-4
   M''(z)^2\right)\right)}{M(z)^4}\nonum
   &&+\ \frac{4 \ln \sX}{(\sX+1) M(z)^4} \Bigl(2 (\sX+1) M(z)^3 M''(z) \ln M(z)+2 (\sX+1) M'(z)^4- (3 \sX+2) M(z)
   M'(z)^2 M''(z)\nonum
   &&\qquad +\  M(z)^2 \left(\sX M''(z)^2-(\sX+1) M'(z)^2 (2 \ln M(z)-1)\right)\Bigr).
\end{eqnarray}
\subsection{Next-to-next-to-leading order}
The NNLO solution is given in the {\it Mathematica} filed attached. 
We only note that $B^{(2)}$ includes the following integration which we could not express by known functions
\begin{eqnarray}
 F_1(x) = \int^\infty_x \frac{\ln ^2t \ln (t+1)-\ln ^3 t}{t(t-1)}dt = \int_0^{1/x} \frac{\ln ^2 u \ln (u+1)}{1-u}du.
\end{eqnarray}
At $x = 1$, this integration converges to $F_1(1)  = 0.3457$.
Fortunately, we do not need this value for the current order analysis.

\section{Next-to-Next-to-Leading order matching}\label{sec:nnlomatch}
We show the matching condition up to $\ord{n^{-2}}$.
The detail of $\cA_0(z)$ is given by
\begin{align}
& \cA_0(z) =4 M(z)+\frac{4 \zeta(2)  M(z)-\frac{4 \zeta(2)  M'(z)^2}{M(z)}+4 (\zeta(2) -\ln M(z)) M''(z)}{n}\nonum
& \quad +\frac{2}{5 n^2 M(z)^3}
   \Bigl[5 M(z)^4 \left(\zeta(2) ^2+4 \zeta (3)\right)-\left(-30 \zeta(2) +\zeta(2) ^2+20 \ln M(z)-5 \zeta
   (3)\right) M'(z)^4\nonum
   &\quad+M(z)^3 \left(20 \zeta(2) +7 \zeta(2) ^2+10 (-1+2 \zeta(2) ) \ln M(z)-15 \ln ^2M(z)+25
   \zeta (3)\right) M''(z)\nonum
   &\quad-M(z) \left(50 \zeta(2) +\zeta(2) ^2-40 \ln M(z)+5 \ln ^2M(z)+10 \zeta
   (3)\right) M'(z)^2 M''(z)\nonum
   &\quad-M(z)^2 \left(\left(10 \zeta(2) +4 \zeta(2) ^2+20 \zeta(2)  \ln M(z)+10 \ln
   ^2M(z)+25 \zeta (3)\right) M'(z)^2\right.\nonum
  &\quad\qquad \left.-\left(20 \zeta(2) +2 \zeta(2) ^2-10 \ln M(z)+10 \ln ^2M(z)+5
   \zeta (3)\right) M''(z)^2\right)\Bigr]. \label{eq:A0-nnlo}
\end{align}
With eq.~(\ref{eq:A0-nnlo}) and $\cB_0(z)$, eq.~(\ref{eq:far-const-0}) becomes
\begin{eqnarray}
&&\beta=\frac{n\cA_0-(n+1)\cB_0}{4n} =M''(z)-\frac{M'(z)^2}{M(z)}+M(z)\nonum
&&+\fr{n}\left[\frac{(4 \ln  2 +1)
   M''(z)^2}{M(z)}+\frac{(4 \ln  2 -\zeta(2) +3) M'(z)^4}{M(z)^3}+\frac{M'(z)^2 (-4 \ln  2 -2
   \ln  M(z)-1)}{M(z)}\right.\nonum
&&\left.+M''(z) \left(4 \ln  2 +\zeta(2) +\frac{(-8 \ln  2 +\zeta(2) -4)
   M'(z)^2}{M(z)^2}+2 \ln  M(z)+1\right)+\zeta(2)  M(z)\right]\nonum
&&+\fr{n^2}\left[\frac{1}{2} M(z) \left(\zeta (2)^2+4 \zeta (3)\right)+\frac{M''(z)^3 \left(16 \ln ^2 2+12 \ln  2 -\frac{8
   \zeta(2) ^2}{5}+\zeta(2) -\frac{\zeta (3)}{2}+1\right)}{M(z)^2}\right.\nonum
   &&+\frac{M'(z)^6 \left(-24   \ln ^2 2-4 \ln  2  (\zeta(2) +6)+\frac{53 \zeta(2) ^2}{10}+4 \zeta(2) +\frac{5 \zeta
   (3)}{2}-7\right)}{M(z)^5}\nonum
   &&\hspace{-1cm} +\frac{M'(z)^2 \left(-16 \ln ^2 2-12 \ln  2  \zeta(2) +\frac{13
   \zeta(2) ^2}{10}+2 \zeta(2) -2 (4 \ln  2 +1) \ln  M(z)-2 \ln  ^2M(z)+\frac{3 \zeta(3)}{2}+2\right)}{M(z)}\nonum
&&\hspace{-1cm}   +\frac{M'(z)^4 \left(40 \ln ^2 2+16 \ln  2  (\zeta(2) +1)-\frac{71
   \zeta(2) ^2}{10}-2 \zeta(2) +4 (4 \ln  2 -\zeta(2) +3) \ln  M(z)-6 \zeta (3)-1\right)}{M(z)^3}\nonum
&&\hspace{-1cm}    +M''(z) \left(8 \ln ^2 2+4 \ln  2  (\zeta (2)+1)+\frac{M'(z)^4
   \left(64 \ln ^2 2 +\ln  2  (8 \zeta(2) +60)-\frac{25 \zeta ^2}{2}-5 \zeta(2) -\frac{11 \zeta
   (3)}{2}+12\right)}{M(z)^4}\right.\nonum
&& \hspace{-1cm}   +\frac{M'(z)^2 \left(-64 \ln ^2 2 -4 \ln  2  (5 \zeta (2)+8)+9
   \zeta(2) ^2-2 \zeta(2) -4 (8 \ln  2 -\zeta(2) +4) \ln  M(z)+\frac{13 \zeta
   (3)}{2}\right)}{M(z)^2}\nonum
   &&\left.+2 (4 \ln  2 +\zeta (2)+1) \ln  M(z)+2 \ln  ^2M(z)+\frac{1}{2}
   \left(\zeta(2) ^2+6 \zeta +4 \zeta(2) (3)-2\right)\right)\nonum
  && +M''(z)^2 \left(\frac{M'(z)^2
   \left(-56 \ln ^2 2 -4 \ln  2  (\zeta(2) +12)+\frac{44 \zeta(2) ^2}{5}+\frac{7 \zeta
   (3)}{2}-6\right)}{M(z)^3}\right.\nonum
&&   \left.\left.+\frac{24 \ln ^2 2 +4 \ln  2  (\zeta(2) +4)-\frac{8 \zeta(2) ^2}{5}+4
   \zeta(2) +4 (4 \ln  2 +1) \ln  M(z)-\frac{\zeta (3)}{2}}{M(z)}\right)\right]
\end{eqnarray}
where the higher derivatives of $M(z)$ are already eliminated by using eq.~(\ref{eq:mom-const-nnlo}).
 The nonlinear terms of $M''(z)$ in the higher order
can be also eliminated by the repeated use of the same equation.
After the integration and again eliminating $M'(z)$ in the higher order, we obtain eq.~(\ref{eq:horizon-eq-2nd}).

\section{Near uniform solution} \label{sec:nnloser}
The solution of eq.~(\ref{eq:horizon-eq-2nd}) expanded with $\tib=1-\epsilon^2/2$ is expressed in the Fourier series
\begin{eqnarray}
 \tiM(z) = \sum_{j=0}^\infty \mu_j(\epsilon) \epsilon^j \cos(j k(\epsilon) z) 
\end{eqnarray}
where the wavenumber $k(\epsilon)$ is given by eq.~(\ref{eq:nnlo-kgl}).
In the following, we show first few terms. The terms up to $\ord{\epsilon^7}$ are given in the attached {\it Mathematica} file. 
\begin{eqnarray}
\mu_0(\epsilon) &=&1-\frac{\epsilon ^4}{96}-\frac{11 \epsilon
   ^6}{3456}
  +\fr{n}\left(-\zeta(2) +\frac{\zeta(2)  \epsilon ^2}{2}+\frac{3 \epsilon
   ^4}{32}+\frac{5 \epsilon ^6}{192}
   \right)\nonum
&&   +\fr{n^2}\left[\frac{\zeta(2) ^2}{2}-2\zeta(3)+\epsilon ^2 \left(\frac{\zeta(2) ^2}{10}-\zeta(2) +\zeta
   (3)+\frac{1}{2}\right)\right.\nonum
&&\left.\qquad
   +\frac{ \left(-91 \zeta(2) ^2+260 \zeta(2) -200\right) \epsilon ^4}{960}+\frac{\left(-497 \zeta(2) ^2+1420 \zeta(2) -1320\right) \epsilon
   ^6}{34560}  \right],\nonum
\end{eqnarray}
\begin{eqnarray}
&&\mu_1(\epsilon)=1-\frac{7 \epsilon ^2}{288}-\frac{341 \epsilon ^4}{82944}-\frac{729281   \epsilon ^6}{597196800}\nonum
   &&+\fr{n}\left(-\frac{1}{2}+2 \ln 2  +\frac{1}{576} (85-124 \ln 2  )
   \epsilon ^2+\frac{(6113-6164 \ln 2  ) \epsilon ^4}{165888}+\frac{(15099791-12711044 \ln 2  ) \epsilon ^6}{1194393600}\right)\nonum
   &&+\fr{n^2}\left[\frac{3}{8}-2 \ln ^2  2-\ln 2   (4 \zeta(2) +1)+\frac{\zeta(2) ^2}{5}+2 \zeta(2)+\frac{\zeta (3)}{2}\right. \nonum
  && +\frac{\epsilon
   ^2 \left(112 \ln ^2  2+8 \ln 2   (508 \zeta(2) +109)-472 \zeta(2) ^2-1264 \zeta(2) -508 \zeta
   (3)-569\right)}{2304}\nonum
    &&+\frac{\epsilon ^4 \left(73360 \ln ^2  2+40 \ln 2   (5780
   \zeta(2) +19697)+50264 \zeta(2) ^2-292240 \zeta(2) -25 (1156 \zeta (3)+8035)\right)}{3317760}\nonum
   &&+\epsilon ^6 \left(\frac{2270081 \ln ^2  2}{298598400}+\left(\frac{1546721 \ln 2 
   }{149299200}-\frac{5865761}{298598400}\right) \zeta(2)\right.\nonum
   &&\qquad \left.\left. +\frac{43223471 \ln 2 
   }{597196800}+\frac{13569919 \zeta(2) ^2}{2985984000}-\frac{11 (562444 \zeta
   (3)+12582813)}{4777574400}\right) \right],
\end{eqnarray}
\begin{eqnarray}
 &&\mu_2(\epsilon)=\frac{1}{6}+\frac{\epsilon ^2}{72}+\frac{8993 \epsilon
   ^4}{2488320}+\fr{n}\left(\frac{1}{3} (4 \ln 2 
   -1)-\frac{\epsilon ^2}{12}+\frac{(-1454 \ln 2  -12997) \epsilon ^4}{622080}\right)\nonum
 &&+\fr{n^2}\left[\frac{1}{12} \left(32 \ln 2  ^2+16 \ln 2   (\zeta(2) -2)-5 \zeta(2) ^2+4 \zeta(2) -2 (\zeta
   (3)+1)\right)\right.\nonum
   &&+\frac{1}{720} \epsilon ^2 \left(-320 \ln ^2  2-80 \ln 2   (4 \zeta(2)
   -5)+107 \zeta(2) ^2-100 \zeta(2) +40 \zeta (3)+110\right)\nonum
   &&\left.+\frac{\epsilon ^4
   \left(-461024 \ln ^2  2+\ln 2   (610496-255856 \zeta(2) )+77435 \zeta(2) ^2-56764 \zeta(2) +31982 \zeta
   (3)-45596\right)}{4976640}\right],\nonum
\end{eqnarray}
\begin{eqnarray}
&&\mu_3(\epsilon)=\frac{1}{96}+\frac{\epsilon ^2}{360}+\frac{10631 \epsilon ^4}{11059200}\nonum
   &&+\fr{n}\left(\frac{3}{64} (4 \ln 2  -3)+\frac{1}{640} (22
   \ln 2  -23) \epsilon ^2+\frac{(24684 \ln 2  -30253) \epsilon ^4}{2457600}\right)\nonum
   &&+\fr{n^2}\left[\frac{1}{768} \left(1008 \ln ^2  2-72 \ln 2   (4 \zeta(2) +11)+104 \zeta(2) ^2-112 \zeta(2) +36 \zeta
   (3)+207\right)\right. \nonum
   \nonum
   &&\quad+\ \frac{\epsilon ^2 \left(1152 \ln ^2  2+18 \ln 2   (52 \zeta(2) -199)-352
   \zeta(2) ^2+404 \zeta(2) -117 \zeta (3)+443\right)}{11520}   \nonum
   &&\quad+\ \epsilon ^4 \left(\frac{3419 \ln ^2  2}{204800}+\left(\frac{1831 \ln 2 
   }{102400}+\frac{39691}{5529600}\right) \zeta(2) -\frac{32419 \ln 2  }{409600}\right.\nonum
 && \hspace{5cm} \left.\left.-\frac{72277 \zeta(2)   ^2}{11059200}+\frac{1916417-197748 \zeta (3)}{88473600}\right)\right].\nonum
\end{eqnarray}

\end{document}